\documentclass[aps,preprintnumbers,amsmath,amssymb,floatfix,showpacs,twocolumn]{revtex4}
\usepackage{epsfig} 
\usepackage{bm}
\usepackage{epsfig,amsbsy,bm,amssymb,amsmath}

\renewcommand{\r}{\mathbf{r}}
\renewcommand{\v}{\mathbf{v}}
  \newcommand{\R}{\mathbf{R}}
\newcommand{\hs}{\hspace*}

\newcommand{\nn}{\nonumber}
  \newcommand{\la}{\langle}
  \newcommand{\ra}{\rangle}

\newcommand{\ba}{\begin{eqnarray}}
\newcommand{\ea}{\end{eqnarray}}
\newcommand{\br}{\begin{eqnarray*}}
\newcommand{\er}{\end{eqnarray*}}
\newcommand{\be}{\begin{equation}}
\newcommand{\ee}{\end{equation}}
\newcommand{\eref}[1] {(\ref{#1})}
\newcommand{\Eref}[1] {Eq.~(\ref{#1})}
\newcommand{\Fref}[1] {Fig. \ref{#1}}

\begin{document}

\title{Time-dependent calculations of transfer ionization by \\
fast proton-helium collision in one-dimensional kinematics}

\author{Vladislav V. Serov}
\affiliation{
Department of Theoretical Physics, Saratov State University, 83
Astrakhanskaya, Saratov 410012, Russia}

\author{ A. S. Kheifets}

\affiliation{Research School of Physical Sciences,
The Australian National University,
Canberra ACT 2601, Australia}

\date{\today}

\begin{abstract}

We analyze a transfer ionization (TI) reaction in the fast
proton-helium collision
$\rm H^+ + He \to H^0 + He^{2+} + e^-$
by solving a time-dependent Schr\"odinger equation (TDSE) under the
classical projectile motion approximation in one-dimensional
kinematics. In addition, we construct various time independent
analogues of our model using lowest order perturbation theory in the
form of the Born series.  By comparing various aspects of the TDSE and
the Born series calculations, we conclude that the recent
discrepancies of experimental and theoretical data may be attributed
to deficiency of the Born models used by other authors. We demonstrate
that the correct Born series for TI should include the momentum space
overlap between the double ionization amplitude and the wave function of the
transferred electron.
\end{abstract}

\pacs{34.70.+e, 34.10.+x, 34.50.Fa}


\maketitle

\section{Introduction}

The transfer ionization (TI) reaction in a fast proton-helium
collision
$\rm H^+ + He \to H^0 + He^{2+} + e^-$
has been studied thoroughly, both experimentally and theoretically,
for a long time. Experimental observation of the fully differential
momentum distribution of the ejected electron
\cite{PhysRevLett.86.2257} revealed the potential of this reaction to
examine radial and angular correlations in the helium atom ground
state. This potential was further explored in
\cite{SB03,Kheifets2003a}. Later on, the experimental set up was
improved to detect the complete three-particle coincidence $\rm
H-He^{2+}-e^-$ and to map three-dimensional ejected electron momentum
distributions \cite{PhysRevA.87.032715,PhysRevA.88.042710,PhysRevA.71.052712}.
On the theoretical side, various calculations were performed employing
the lowest order perturbation theory in the form of the Born series.
The initial first Born results
\cite{0953-4075-37-10-L01,PhysRevA.71.052712} confirmed that the fully
differential cross sections for TI were indeed a sensitive probe of
the ground state correlation in the helium atom ground state. By
employing ground state wave functions of various level of
sophistication, better or worse agreement with the experiment could be
achieved.  A more detailed comparison with experimental data, in the
form of the fully differential cross sections, was not possible at the
time. Indeed, in the earlier experiments \cite{PhysRevLett.86.2257,
SB03}, the momentum distribution of the ionized
electron was derived from the momentum and energy conservation with
other collision partners, but not directly measured as in the latest
experiments
\cite{PhysRevA.87.032715,PhysRevA.88.042710}. Nevertheless, the
authors of Ref.~\cite{0953-4075-37-10-L01} continued their quest and
included the second Born corrections into their model in the form of
the closure term \cite{PhysRevA.78.012714}. Improvement to the first
Born results was marginal in terms of their agreement with the
experimental data.

A similar first Born model of another group of authors
\cite{PhysRevA.81.032703} was used to interpret three-dimensional
electron-momentum distributions in a series of joint experimental and
theoretical works \cite{PhysRevA.87.032715,PhysRevA.88.042710}.
Similarly to the initial studies by Godunov {\em et al}
\cite{0953-4075-37-10-L01,PhysRevA.71.052712,PhysRevA.78.012714},
a strong sensitivity of the calculated results was demonstrated to the
quality of the helium atom ground state wave function. However, agreement
with the experimental data was qualitative at best. It was hoped that
extension of the theoretical model to the second Born treatment would
have improved this agreement. However, such an extension reported
earlier \cite{PhysRevA.78.012714} was not very efficient.  General
utility of the second Born corrections is discussed in \cite{0953-4075-32-15-104}.

In their perturbative treatment, both groups of authors identified
three terms in the first Born amplitude of the TI reaction.  These
terms can be associated with various terms of the interaction
potential
$
V_{p1}+V_{p2}+V_{Np}
$
when this amplitude is expressed in its {\em prior} form. This
potential describes the interaction of the projectile proton with
the two target electrons and the nucleus.

The first term, associated with $V_{p1}$ and denoted by $A_1$ in
\cite{PhysRevA.81.032703,PhysRevA.87.032715,PhysRevA.88.042710} or
termed ``transfer first'' in
\cite{0953-4075-37-10-L01,PhysRevA.71.052712,PhysRevA.78.012714}, is
shown to be identical to the Oppenheimer, Brinkmann, and Kramers (OBK)
amplitude and describes the collision between the target electron
labeled 1 and the proton followed by the capture of this electron by
the projectile. The second electron, labeled 2, is released due to
rearrangement in the helium atom, known as the shake-off (SO)
process. In principle, in the $\rm e - H^+$ interaction can be
factored out from the OBK term $A_1$ \cite{PhysRevA.87.032715}. The
remaining SO amplitude provides a simplified theoretical treatment of
TI that was applied in \cite{SB03,Kheifets2003a}.
The second term, associated with $V_{p2}$ and denoted by $A_2$ or termed
as ``ionization first'', describes the process in which the proton
knocks off a target electron 1  into the continuum, followed by
capturing the remaining electron 2 from the helium atom. 
The third term, associated with $V_{pN}$ and denoted by $A_3$
represents an initial interaction between the
projectile and the helium nucleus followed by the electron
capture. Similarly to $A_1$, the second electron is also released due
to the sudden rearrangement in the helium ion.

Unlike the other authors, Voitkiv {\em et al}
\cite{PhysRevLett.101.223201,0953-4075-41-19-195201,PhysRevA.86.012709}
expressed  the first Born amplitude in the {\em post} form
which contained a different interaction potential
$V_{N1}+V_{p2}+V_{12}+V_{Np} \ .$
They identified an additional electron-electron Auger mechanism of TI
associated with the term $V_{12}$. In this mechanism, the electron to
be transferred rids itself of the excess energy not via the coupling
to the radiation field, as in the radiative capture process, but 
by  interaction with the other electron.  In their comment on
\citet{PhysRevA.86.012709}, \citet{PhysRevA.89.036701} argued that
since both the {\em prior} and {\em post} forms of the first Born
amplitude should be identical, the newly discovered electron-electron
Auger mechanism was, in fact, contained in the long-known OBK
amplitude. In their reply, \citet{PhysRevA.89.036702} retorted this
argument by appealing to the physical intuition and analogy between
the TI and other related processes. Other than this appeal, no
quantitative arguments were provided.

Outside the first Born treatment lies the process of repeated
interaction of the projectile with the target in which both the target
electrons are ejected in sequence. This sequential process is known in
the electron-impact ionization and double photoionization as the
two-step-2 process \cite{McGuire1997}. For the purpose of the present
study we will call it binary encounter (BE). The signatures of the SO
and BE processes in TI can be found in the momentum distribution of
the ejected electron.  In the BE process, the ejected electron flies
predominantly in the forward direction (in the direction of the
projectile), due to the momentum transferred from the projectile.  In
the SO process, the emitted electron flies predominantly backwards
because the instantaneous momenta of an electron pair in the helium
atom ground state are aligned in the opposite directions. The SO
process is contained in the first Born amplitude while the BE process
can be only accommodated by further terms in the Born series.  At
small velocities of the projectile, the BE mechanism dominates whereas
at large velocities it is the SO mechanism that becomes dominant. 

In the absence of a rigorous theoretical treatment, when the Born
series calculations cannot reproduce the experimental data on a
quantitative level, an additional insight into the TI reaction can be
gained by a non-perturbative approach based on solving the
time-dependent Schr\"odinger equation (TDSE).  For a four-body Coulomb
problem like TI, the TDSE cannot be solved in its full dimensionality
and additional approximations should be made.
In the experiments
\cite{PhysRevLett.86.2257,PhysRevA.87.032715,PhysRevA.88.042710}, the
projectile proton had the velocity $v_p\simeq 5$~a.u. and the momentum
$p_p\simeq 10^4$~a.u. This corresponds to the wavelength
$\lambda_p\sim 10^{-3}$~a.u. which is much less then the atom
size. Due to this fact and because of a small relative change of the
projectile velocity, the classical projectile motion approximation
(CPA) should be sufficiently accurate. The TI reaction in this
approximation is described by a six-dimensional TDSE which can be
solved, in principle, using modern computational facilities.
Nevertheless, for the purpose of the present work, we further simplify
the problem and restrict the motion of all the particles to one
dimension (1D).  We compare results of thus restricted TDSE
calculation with various perturbative Born calculations, also reduced
to 1D.  By analyzing various aspects of the TDSE and the Born series
calculations, we conclude that the recent discrepancies of
experimental and theoretical data may be explained by deficiency of
theoretical approach used by other authors. We demonstrate that the
correct Born series for TI should include the momentum space overlap
between the ionization amplitude and the wave function of the
transferred electron.

\section{Theoretical model} 

Within the scope of the CPA, the full-dimensional TDSE takes the form
\ba
i\frac{\partial\Psi(\r_1,\r_2,t)}{\partial  t}&=&
\left[\hat{H}_0-\frac{1}{|\r_1-\R(t)|}-\frac{1}{|\r_2-\R(t)|}\right]
\nn\\ &&\hs{0.5cm}\times\Psi(\r_1,\r_2,t) \ ,
\label{TDSE6D} 
\ea 
with the initial condition 
\be
\Psi(\r_1,\r_2,t_0)=\Phi_0(\r_1,\r_2)\exp(-iE_0t_0);
\quad t_0\to -\infty.  
\ee 
Here $\R(t)=(b,0,v_p t)$ is a current position of a proton, $b$ is an impact parameter, $\Phi_0(\r_1,\r_2)$ is the target ground state function, and
$\hat{H}_0$ is target Hamiltonian, which for the helium atom has the
form
\ba
\hat{H}_0=
-\frac12\nabla_1^2 -\frac12\nabla_2^2
-\frac{1}{r_1} -\frac{1}{r_2}+\frac{1}{|\r_2-\r_1|}.
\ea
For a one-dimensional kinematics, \Eref{TDSE6D} is reduced to 
two-dimensional TDSE 
\ba 
i\frac{\partial\psi(x_1,x_2,t)}
{\partial   t} &=&
\left[\hat{H}_0+U_1(x_1-v_pt)+U_1(x_2-v_pt)\right]
\nn\\&&\hs{5mm}\times\psi(x_1,x_2,t)
\label{TDSE2D} 
\ea 
with the initial condition 
\be
\psi(x_1,x_2,t_0)=\varphi_0(x_1,x_2)\exp(-iE_0t_0); \quad t_0\to
-\infty 
\ee 
and the Hamiltonian 
\ba
\hat{H}_0&=&
-\frac{1}{2}\frac{\partial^2}{\partial x_1^2}
-\frac{1}{2}\frac{\partial^2}{\partial x_2^2}
+U_2(x_1) +U_2(x_2)
\nn\\&&\hs{3cm}
+ U_{-1}(x_2-x_1).  
\ea 
Here the effective potentials $U_Z(x\to\pm\infty)=-Z/|x|$
is taken in the form of  a shifted Coulomb potential 
\be U_Z(x)=-\frac{Z}{|x|+|Z|^{-1}} \ .
\label{shiftCoul} 
\ee 
With this potential, the He$^+$ ion is described by a  1D
equation
\ba 
\left[-\frac{1}{2}
\frac{\partial^2}{\partial x^2}
+U_Z(x)\right]\varphi^Z_{n}(x)=
\epsilon^Z_{n}\varphi^Z_{n}(x)
\ea 
with the ground state energy $\epsilon^Z_{0}=-Z^2/2$ being equal to
the ground state energy of a conventional 3D ion. The ground state
wave function
\be
\varphi^Z_{0}(x)=\sqrt{\frac{2Z}{5}}(1+Z|x|)\exp(-Z|x|) 
\ ,
\ee
when Fourier transformed to the momentum space 
\be 
\varphi^Z_{0}(x)=\frac{1}{\sqrt{2\pi}}\int_{-\infty}^\infty
u^Z_0(\varkappa)e^{i\varkappa x}d\varkappa \ ,
\ee
becomes
\be
u^Z_0(p)=\frac{4}{\sqrt{5\pi Z}}\frac{1}{[1+(p/Z)^2]^{2}} \ .  
\ee 
This only differs by a normalization constant from a conventional
ground state wave function of a hydrogenic ion in the momentum space
$$
u_{100}(p)= \frac{2\sqrt{2}}{\pi
  Z^{3/2}}\frac{1}{[1+(p/Z)^2]^{2}} \ .
$$
It is a useful property that allows one to maintain the correct
dependence of the TI cross-section on $v_p$ which is determined by the
momentum space overlap.
The Fourier transform of the potential \eref{shiftCoul}
\ba 
V_Z(q)&=&\int_{-\infty}^\infty e^{-iq\xi}
U_Z(\xi) d\xi, 
\\
&=&\nn
2Z\left(\sin\left|\frac{q}{Z}\right|
\mathrm{Ssi}\left|\frac{q}{Z}\right|+
\cos\left|\frac{q}{Z}\right|
\mathrm{Ci}\left|\frac{q}{Z}\right|\right),
\ea 
can be expressed via the cosine integral $\mathrm{Ci}\,(x)$ and
the shifted sine integral $\mathrm{Ssi}\,(x)$

The wave function of the electron captured by the projectile is
\be
\psi_\text{tr}(x,t)=\varphi^1_{0}(x-v_pt)
\exp\left[iv_px-i\left(\frac{v_p^2}{2}+
\epsilon^1_{0}\right)t\right].
\label{psitr}
\ee 
If the second electron is ejected with the momentum $k$, the
two-electron wave function of the final state can be written as
\ba 
\psi_{k\text{H}}(x_1,x_2,t) &=&
\frac{1}{\sqrt{2}}\times
\\ &&\hs{-3cm}
\left[\psi_\text{tr}(x_1,t)\varphi_k^{2(-)}(x_2)+\psi_\text{tr}(x_2,t)\varphi_k^{2(-)}(x_1)\right]e^{-i\frac{k^2}{2}t}
\ ,
\nn
\ea
where $\varphi^{Z(-)}_{k}(x)$ is the continuum state function for the
ejected electron with the energy $\epsilon=k^2/2$.  The amplitude of
TI is given by the following expression
\be
A(k)=\lim_{t\to\infty}\int_{-\infty}^\infty\int_{-\infty}^\infty
\psi_{k\text{H}}^*(x_1,x_2,t)\psi(x_1,x_2,t)dx_1dx_2.  
\label{Aofk}
\ee
In the 1D case, the role of differential cross-section of TI is
assumed by the probability density
\be 
P^{(1)}(k)=\frac{dP}{dk}=|A(k)|^2
\ .
\label{Pdens}
\ee

Using the exchange symmetry of the two-electron wave function
$\psi(x_2,x_1,t)=\psi(x_1,x_2,t)$ , we can split the integration in
\Eref{Aofk} in two steps.  First, we calculate the wave function of
the second electron  when the first electron is transferred
\be 
\chi(x_2,t)=\sqrt{2}\int_{-\infty}^\infty
\psi_\text{tr}^*(x_1,t)\psi(x_1,x_2,t)dx_2
\ . 
\ee
Next, we  calculate the
amplitude of ejection of the second electron
\be
A(k)=\lim_{t\to\infty}\int\limits_{-\infty}^\infty
\varphi^{2(-)*}_{k}(x)\exp\left(i\frac{k^2}{2}t\right)\chi(x,t)dx
\ .
\ee
In the second step, for extraction of the ionization amplitudes from
$\chi(x_2,t)$, we used the {\sc e-surff} method \cite{PhysRevA.88.043403}.
\Eref{TDSE2D} was solved numerically using the simplest 3-point finite
difference scheme for evaluation of the space derivatives, and the
split-step method for the time evolution. The target ground state
function $\varphi_0(x_1,x_2)$ was calculated using the evolution in
the imaginary time providing the ground state energy  $E_0=-3.35$~a.u.

To compare the role of the SO and BE processes, we also performed
calculations for the case of zero inter-electron potential. In such a
case, the SO is absent and only the BE contributes to TI. For that
reason we named this approximation BECPA. In this approximation,
\Eref{TDSE2D}  can be split in two identical equations
\ba
i\frac{\partial\psi(x,t)}{\partial
t}=\left[\hat{h}_0+U_1(x-v_pt)\right]\psi(x,t) 
\ea 
with 
\ba
\hat{h}_0=-\frac{1}{2}\frac{\partial^2}{\partial x^2}+U_2(x) 
\ea 
and the initial condition
\be
\psi(x,t_0)=\varphi^2_{0}(x)e^{-i\epsilon^2_{0}t_0}; \quad t_0\to
-\infty.  
\ee
The wave function of a non-transferred electron in this
approximation is
\be
\chi(x,t)=\sqrt{2}
C_\text{tr}[\psi(x,t)-C_\text{tr}\psi_\text{tr}(x,t)]
\ ,
\ee
where the transfer amplitude is
\be
C_\text{tr}=\lim\limits_{t\to\infty}\int_{-\infty}^\infty
\psi_\text{tr}^*(x,t)\psi(x,t)dx.  
\ee

\section{Results and discussion}

\begin{figure}[ht]
\includegraphics[angle=-90,width=0.85\columnwidth]{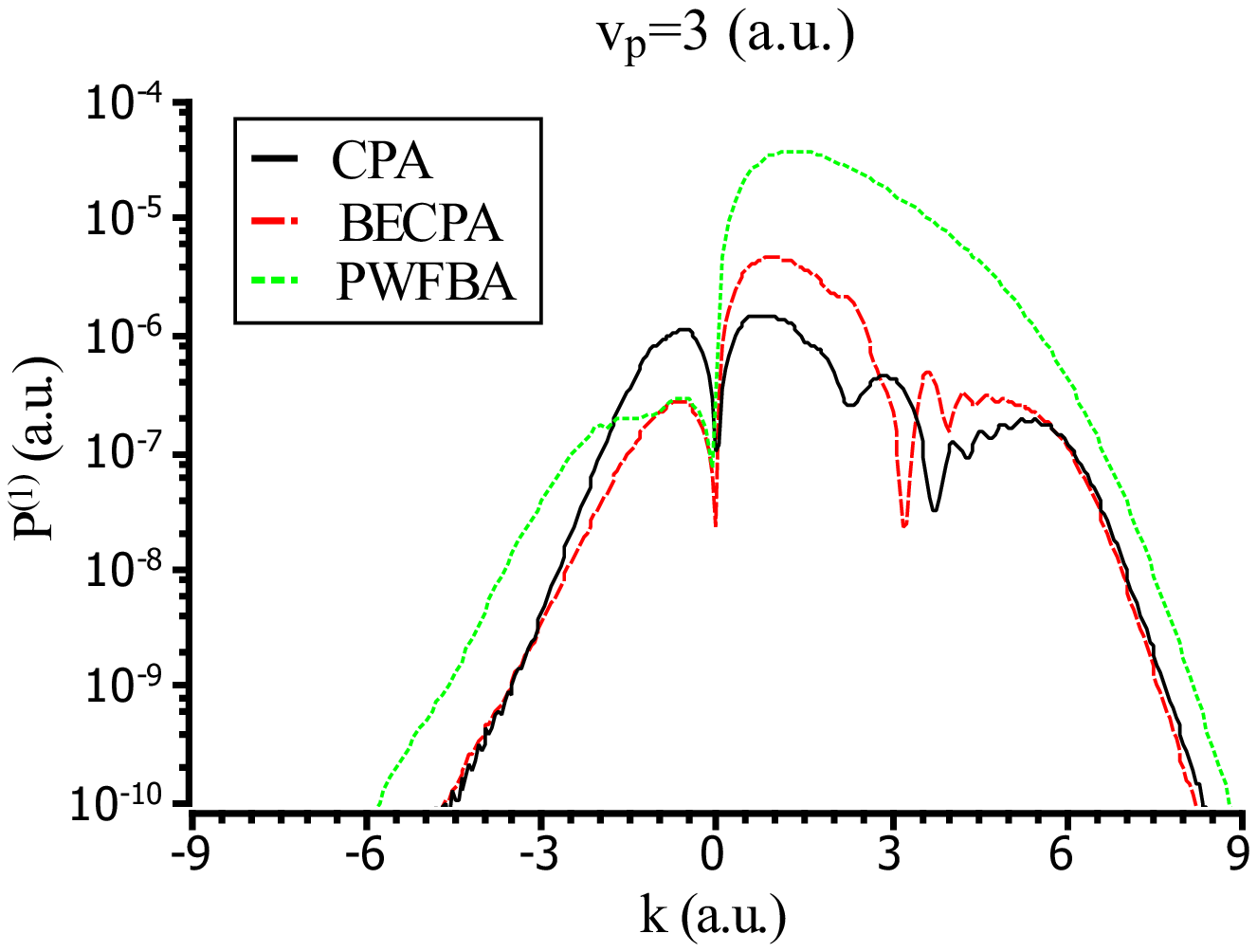}
\includegraphics[angle=-90,width=0.85\columnwidth]{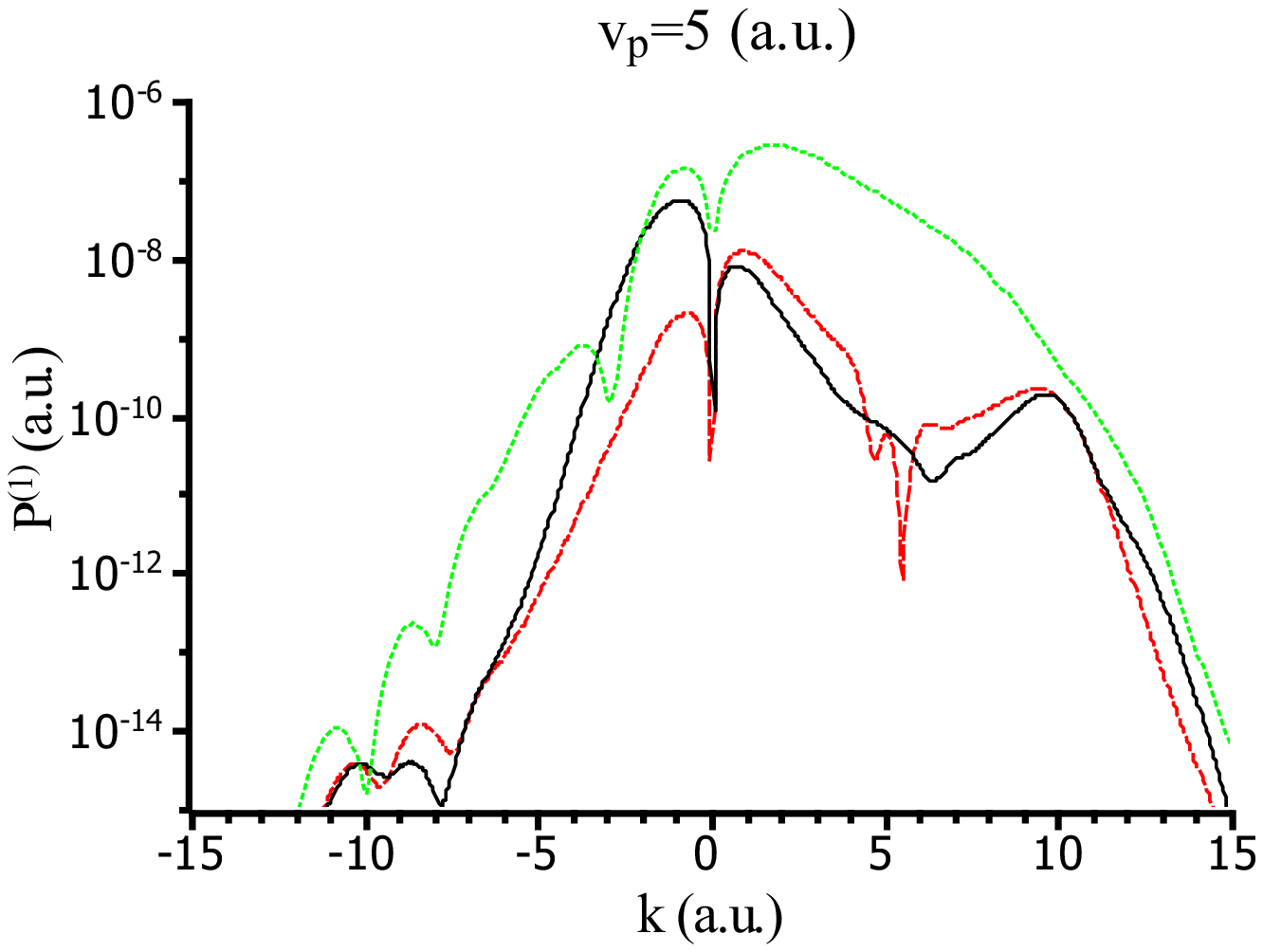}
\includegraphics[angle=-90,width=0.85\columnwidth]{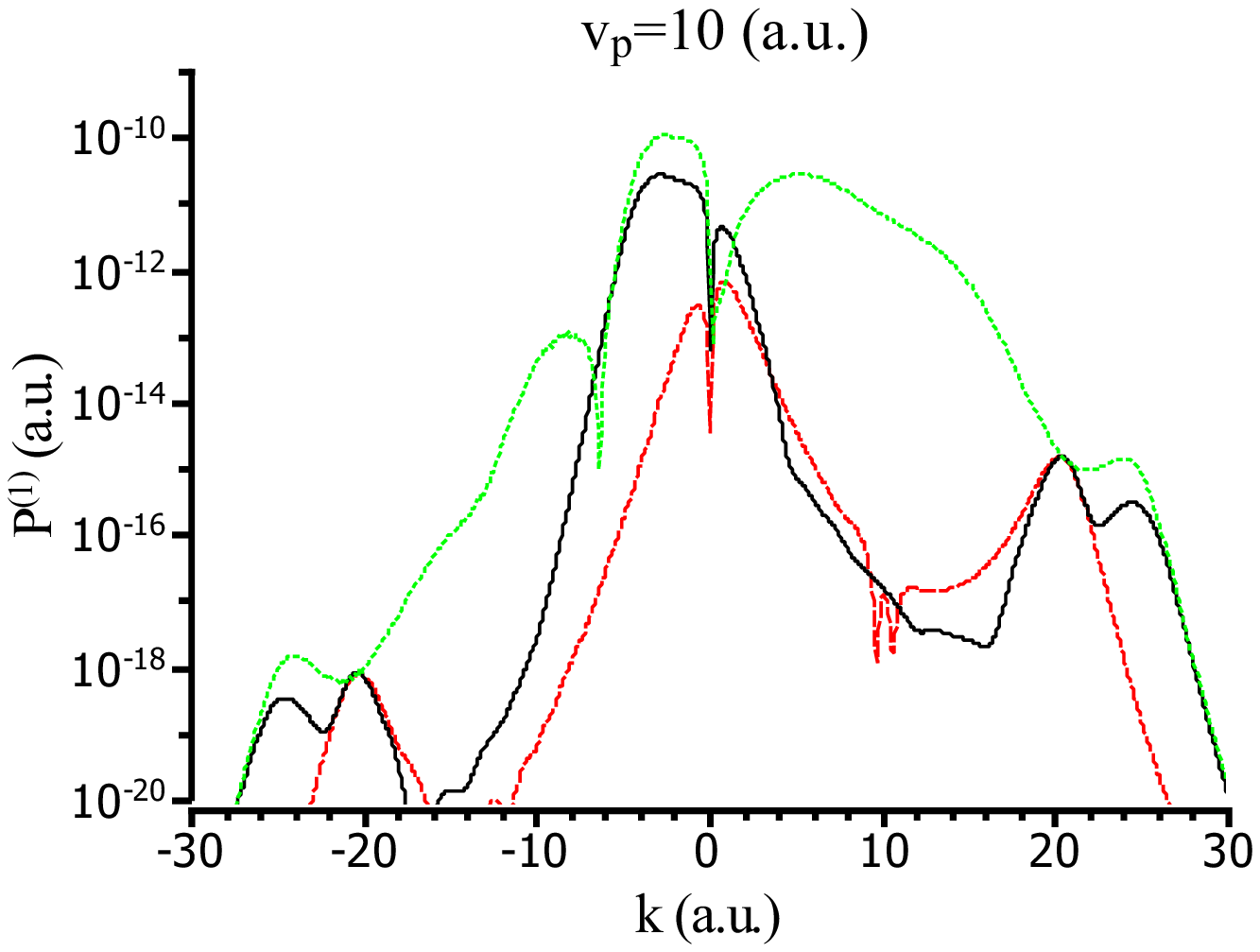}
\caption{(Color online) The TI probability density $P^{(1)}(k)$ as a
function of the ejected electron momentum for the proton velocities
$v_p=3$~a.u. (top), 5~a.u. (middle) and 10~a.u. (bottom). Various
calculations are displayed with the following line styles: the full
CPA (black solid line), the BECPA (red dashed line) and the PWFBA
(green dotted line).}
\label{FIG_P_vs_pe}
\end{figure}

The probability density $P^{(1)}(k)$ \eref{Pdens} as a function of the
ejected electron momentum is shown in \Fref{FIG_P_vs_pe} for the three
selected proton velocities $v_p=3$~a.u. (top), 5~a.u. (middle) and
10~a.u. (bottom). Results of the full CPA and BECPA are shown with the
black solid line and red dashed line, respectively.  The probability
density displays two peaks near $k=0$ and $k=2v_p$. The origin of the
second peak is clear if we consider TI in the rest frame of the
projectile. In this frame, the initial electron wave packet is
scattered on the proton and part of it is reflected back. This
reflected part of the wavepacket has velocity $2v_p$ in the laboratory
frame. The peak near $k=0$ is splitted by the kinematic node at $k=0$.
This node is a characteristic feature for 1D systems with attractive
potentials.

\begin{figure}[ht]
\includegraphics[angle=-90,width=0.85\columnwidth]{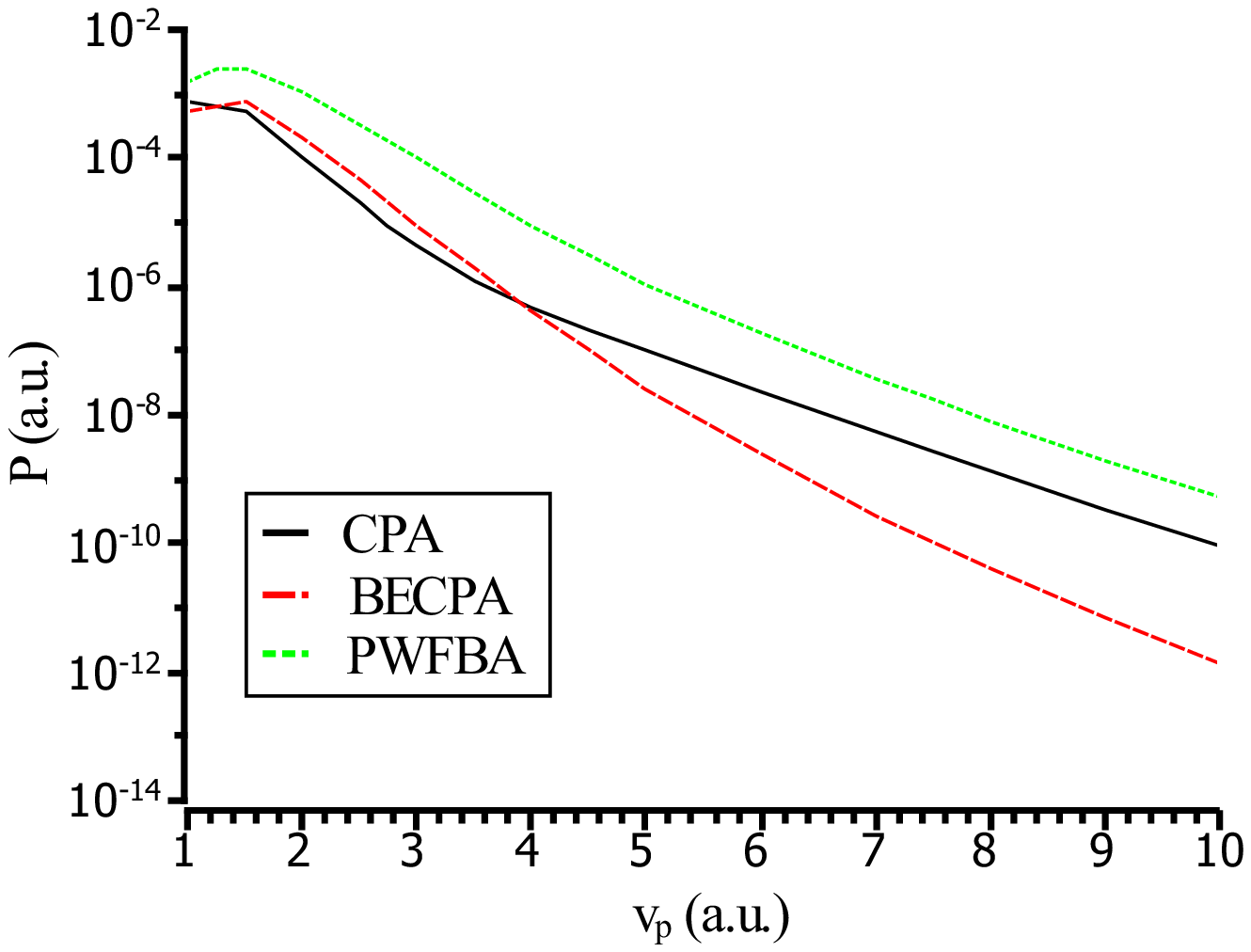}
\includegraphics[angle=-90,width=0.85\columnwidth]{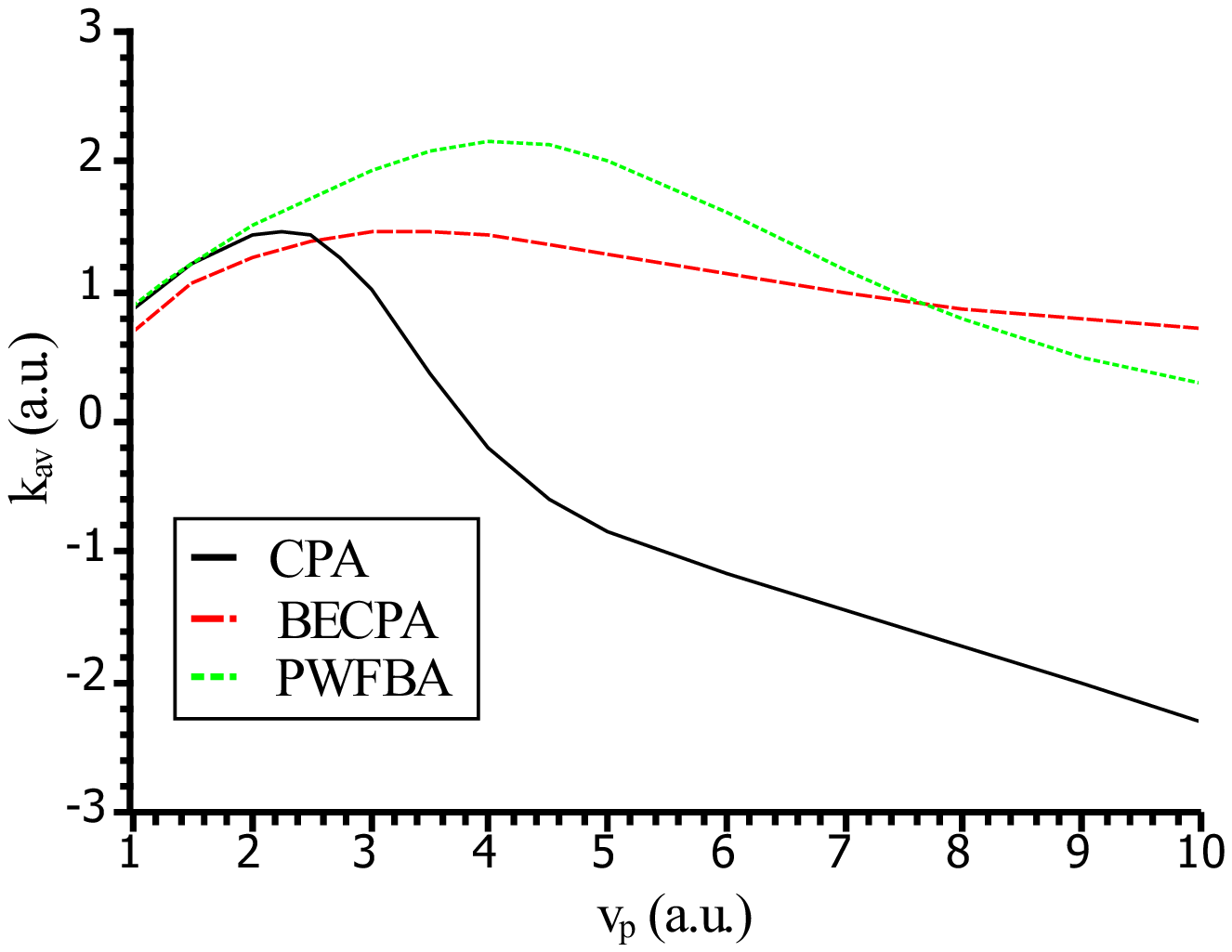}
\caption{(Color online) The TI probability $P$ (top) and the mean
  momentum of ejected electrons $\la k\ra$ (bottom) as
  functions of the projectile velocity $v_p$. Various calculations are
  shown as CPA (black solid line), BECPA (red dashed line), PWFBA
  (green dotted line).}
\label{FIG_k_vs_pe}
\end{figure}

The most essential feature of the experimental differential
cross-sections reported by Sch\"offler {\em et al}
\cite{PhysRevA.87.032715,PhysRevA.88.042710} is the shift of the
maximum of the ejected electron momentum distribution in the direction
opposite to the projectile motion. In the meantime, the 1st Born
cross-sections, reported in the same works, display the main peak
which is largely centered around the zero momentum. 

It is clearly seen in \Fref{FIG_P_vs_pe} that the CPA results
demonstrate the same backward shift of the main peak, except for the
case of the smallest $v_p=3$~a.u. The BECPA results demonstrate the
forward shift for all $v_p$. It is easy to explain this behavior by
the following qualitative arguments. In the BE process, both the
electrons are pushed by projectile, and ionization occurs irrespective
of transfer. Since the momentum transferred from the projectile is
directed forward, main peak in the ionization probability is also
shifted forward. In the SO process, the second electron is preferably
ejected in the direction opposite to the first electron motion which
is captured by projectile. The resulting preferable ejection direction
depends on the relative weighting of the BE and SO process. Since the
SO appears in the first term of the Born series over the
projectile-target interaction, and the BE can only be accommodated by
the second and further terms, the SO is dominant at larger $v_p$,
where the ejected electron should be emitted preferentially in the
direction opposite to the projectile motion.

To compare the relative contributions of the BE and SO processes and
the direction of the preferred emission of the ejected electron
depending on $v_p$, we calculated the total probability of TI and the
mean momentum of the ejected electron
\ba 
P &=& \int_{-\infty}^\infty P^{(1)}(k) dk
\\
\la k\ra &=& \frac{1}{P}\int_{-\infty}^\infty k P^{(1)}(k) dk.
\ea 
By comparing the CPA and BECPA results shown in \Fref{FIG_k_vs_pe}, it
is clearly seen the the BE is dominant at $v_p<4$ while for larger
$v_p$ it is the SO that dominates. In about the same momentum range,
the mean momentum $\la k\ra$ changes its sign. At larger
$v_p$, the mean momentum becomes large and negative. This indicates
that the present CPA results  are consistent with the
experimental observations. One may suggest that in the first Born
approximation (FBA) this preferred backward emission would be even
more prominent feature. However, the plane wave first Born
approximation (PWFBA) \cite{PhysRevA.81.032703}
strongly overestimates forward emission. 

This implies that the deviations of the PWFBA calculations from the
experiment \cite{PhysRevA.87.032715,PhysRevA.88.042710} may be
attributed to the shortcomings of this specific implementation rather
than contribution from the further Born terms.

\subsection{Plane-wave first-Born approximation} 

Let us construct a 1D analogue of the PWFBA.
In this approximation, the transitional amplitude takes the form
\be 
A(k) = \la p_H f |U_1(x_1-x_p)+U_1(x_2-x_p)-2U_1(x_p)| 
p_p 0 \ra \ , 
\label{PWFBA} 
\ee
where the initial state wave function
\ba
\la x_1, x_2, x_p| p_p 0 \ra &=&
\frac{1}{\sqrt{v_p}}e^{ip_px_p}\varphi_0(x_1,x_2) \ .  
\ea 
Following  \cite{PhysRevA.81.032703}, we express the final state wave
function in its asymptotic form
\ba 
 \la x_1, x_2, x_p| p_H f \ra &=&
\frac{1}{\sqrt{2v_H}}e^{ip_fx_p} \times
\\ &&\hs{-3.5cm}
\left[e^{iv_Hx_1}\varphi^1_{0}(x_1-x_p)\varphi_k^{2(-)}(x_2)+
e^{iv_Hx_2}\varphi^1_{0}(x_2-x_p)\varphi_k^{2(-)}(x_1)\right]
\nn\ .
\ea
Here $v_H$ is the velocity of the neutral hydrogen atom, $p_f=mv_H$ is
the proton momentum in the final state. The continuum state wave
functions are normalized by the 1D factors $v_p^{-1/2}$ and
$v_H^{-1/2}$. The continuum normalization for various dimensionality
is addressed in \cite{LL85}.

The momentum transfer from the projectile to the target can be
expressed as 
\be q=p_p-p_f=m(v_p-v_H)\simeq
\frac{1}{v_p}\left[\frac{v_p^2}{2}+\epsilon^1_0+\frac{k^2}{2}-E_0\right].
\ee
Similarly to \cite{PhysRevA.81.032703}, we introduce the momentum
difference of the proton projectile and the neutral hydrogen atom.
\be Q=p_H-p_p =
(m+1)v_H-mv_p = v_H-q. 
\ee
We note that in \cite{PhysRevA.81.032703} this quantity is denoted by
$q$ which is reserved in the present work to $q=v_H-Q$. Following the
cited paper, we split \Eref{PWFBA} into the three distinct terms and
express them by using the Fourier transform of the functions
$\varphi^{1*}_{0}(x-x_p)$ and $U_1(x)$:
\ba
 &&A_1(k)
=\sqrt{2}
\iiint\limits_{-\infty}^{\hs{3mm} +\infty}dx_1dx_2dx_p
e^{iqx_p-iv_Hx_1} \label{A1}
\\&\times&
\varphi^{1*}_{0}(x_1-x_p)\varphi_k^{2(-)*}(x_2)U_1(x_1-x_p)
\varphi_0(x_1,x_2)
\nn\\ &&=\int_{-\infty}^\infty
u^{1*}_0(\varkappa)V_1(q-\varkappa) I_k(v_H-q,0) d\varkappa 
\nn
\ea
\ba
&&A_2(k)
=\sqrt{2}
\iiint\limits_{-\infty}^{\hs{3mm} +\infty}
dx_1dx_2dx_p \label{A2}
e^{iqx_p-iv_Hx_1}
\\&\times&
\varphi^{1*}_{0}(x_1-x_p)\varphi_k^{2(-)*}(x_2)U_1(x_2-x_p)
\varphi_0(x_1,x_2)
\nonumber\\ &&=\int_{-\infty}^\infty
u^{1*}_0(\varkappa)V_1(q-\varkappa) I_k(v_H-\varkappa,-q+\varkappa)
d\varkappa 
\nn
\ea
\ba
 &&A_3(k)
=-2\sqrt{2}
\iiint\limits_{-\infty}^{\hs{3mm} +\infty}
dx_1dx_2dx_p \label{A3}
e^{iqx_p-iv_Hx_1}
\\&\times&
\varphi^{1*}_{0}(x_1-x_p)\varphi_k^{2(-)*}(x_2)U_1(x_p)
\varphi_0(x_1,x_2)
\nonumber\\ &&=-2\int_{-\infty}^\infty
u^{1*}_0(\varkappa)V_1(q-\varkappa)I_k(v_H-\varkappa,0) d\varkappa 
\nn
\ea
Here we introduced the notation
\ba
I_k(\kappa_1,\kappa_2)&=&\frac{\sqrt{2}}{(2\pi)^{1/2}}
\int_{-\infty}^\infty
\varphi_k^{2(-)*}(x_2)e^{-i\kappa_2x_2}
\\&\times&
\left[\int_{-\infty}^\infty
e^{-i\kappa_1x_1} \varphi_0(x_1,x_2)dx_1\right]dx_2.  
\nn
\ea
Finally, \Eref{PWFBA} takes the form
\ba
A(k)&=&\frac{1}{\sqrt{v_pv_H}}\int_{-\infty}^\infty
d\varkappa 
u^{1*}_0(\varkappa)V_1(v_H-Q-\varkappa)
\\&&
\hs{-1.5cm}
\left[ I_k(Q,0) +
I_k(v_H-\varkappa,-v_H+Q+\varkappa) -2I_k(v_H-\varkappa,0)\right]
\nn
\label{PWFBAfin} 
\ea
Which, apart from notations, coincides with Eq.(2) of
\citet{PhysRevA.81.032703} 

In \Fref{FIG_P_vs_pe} we display the probability density calculated
with the PWFBA. It is clear that this calculation deviates strongly
from the non-perturbative CPA calculation. In the PWFBA, the
probability density displays a peak at $k>0$ which overshoots strongly
the CPA peak, both by the magnitude and the width.  As is seen on the
bottom panel of \Fref{FIG_k_vs_pe}, this overestimation leads to the
mean momentum $\la k\ra > 0$ even at large $v_p$. Hence, the
1D implementation of the PWFBA displays the same characteristic
feature as the original implementation used in
\cite{PhysRevA.87.032715,PhysRevA.88.042710}. 

In order to elucidate the origin of this behaviour, we express the
PWFBA amplitude neglecting the inter-electron interaction.
In this case,
$\varphi_0(x_1,x_2)=\varphi^2_{0}(x_1)\varphi^2_{0}(x_2)$ 
and
\br
I_k(\kappa_1,\kappa_2)&=&
\frac{\sqrt{2}}{(2\pi)^{1/2}}\int_{-\infty}^\infty
\varphi_k^{2(-)*}(x_2)
e^{-i\kappa_2x_2}\varphi^2_{0}(x_2)dx_2
\\&&\times
\int_{-\infty}^\infty
e^{-i\kappa_1x_1} \varphi^2_{0}(x_1)dx_1
=\sqrt{2}u^2_{0}(\kappa_1)I_k(\kappa_2),
\er
where
$$
 I_k(\kappa)=\int_{-\infty}^\infty \varphi_k^{2(-)*}(x)
e^{-i\kappa x}\varphi^2_{0}(x)dx \ .
$$
Then \Eref{PWFBAfin} takes the form
\br
 A(k)&=&\frac{\sqrt{2}}{\sqrt{v_pv_H}}\int_{-\infty}^\infty
u^{1*}_0(\varkappa-v_H)V_1(\varkappa-Q)
\\
&&\hs{-1cm}
\left[ u^2_{0}(Q)I_k(0) +
u^2_{0}(\varkappa)I_k(Q-\varkappa) -2u^2_{0}(\varkappa)I_k(0)\right]
d\varkappa \ .
\er
Since $I_k(0)=0$, the only second term survives under the integral
sign, i.e.
\br
A(k)&=&A_2(k)\\
&&\hs{-1.5cm}=
\frac{\sqrt{2}}{\sqrt{v_pv_H}}\int_{-\infty}^\infty
u^{1*}_0(\varkappa-v_H)V_1(\varkappa-Q)I_k(Q-\varkappa)u^2_{0}
(\varkappa)d\varkappa.
\er
The authors of
Ref.~\cite{PhysRevA.81.032703,PhysRevA.87.032715,PhysRevA.88.042710}
claim that this term is responsible for the BE process. However, BE
should be zero in the FBA. The projectile can only act on one of the
target electrons and the second electron makes no transition in the
absence of the inter-electron interaction. The reason why the term
$A_2$ is not zero becomes clear if we write it in its original form
as the coordinate integral
\br
A_2(k) &=& \sqrt{2}\int_{-\infty}^\infty \hs{-4mm} dx_p e^{iqx_p}
\int_{-\infty}^\infty \hs{-4mm} dx_1 e^{-iv_Hx_1}
\varphi^{1*}_{0}(x_1-x_p)\varphi^2_0(x_1)
\\&&\hs{.5cm}\times
\int_{-\infty}^\infty
\varphi_k^{2(-)*}(x_2)U_1(x_2-x_p)
\varphi^2_0(x_2)dx_2 \ .
\er
One can see that the projectile interaction causes ionization (the integral
over $x_2$) and the transfer takes place due to the non-orthogonality
of the initial and final state wave functions
$$
\la\varphi^2_0(x)|\,e^{iv_Hx}\varphi^{1}_{0}(x-x_p)\ra 
\neq 0
$$
Thus $A_2$ is the artifact of the approximation employed in 
\cite{PhysRevA.81.032703,PhysRevA.87.032715,PhysRevA.88.042710}.

Note that the term $A_3$ appears due to proton-nucleus interaction. In
CPA, the proton-nucleus potential only adds an overall time-dependent
phase factor to the wave function, and thus is unable to produce
change of the electronic state. In the Jackson--Schiff (JS) theory of a
related process of simple transfer, an introduction of the
proton-nucleus potential compensates a spurious term appearing due to
the non-orthogonality of the initial and final states
\cite{Bates1952,Bates1958}. For transfer ionization, $A_3$ plays an
analogous role, but it does not provide the full compensation.

It is possible to eliminate the spurious electron transfer by
orthogonalizing the single-particle initial and captured electron
states. For this purpose, in Eqs.~\eref{A1}, \eref{A2} and \eref{A1}
one shall replace $\varphi_0(x_1,x_2)$  by
\ba 
&&\tilde{\varphi}_0(x_1,x_2,x_p) \equiv \varphi_0(x_1,x_2)
\nn\\&&\hs{.5cm}
 - \la e^{iv_Hx_1}\varphi^1_{0}(x_1-x_p)|
\varphi_0(x_1,x_2) \ra e^{iv_Hx_1}\varphi^1_{0}(x_1-x_p)
\nn\ .
\ea
It is easy to see that $A_2\equiv 0$ and $A_3\equiv 0$ after this
replacement. However, this operation does not assure that $A_1$ gives
the correct result.

\subsection{Other theoretical models of TI in 1D }

\begin{figure}[ht]
\includegraphics[angle=-90,width=0.85\columnwidth]{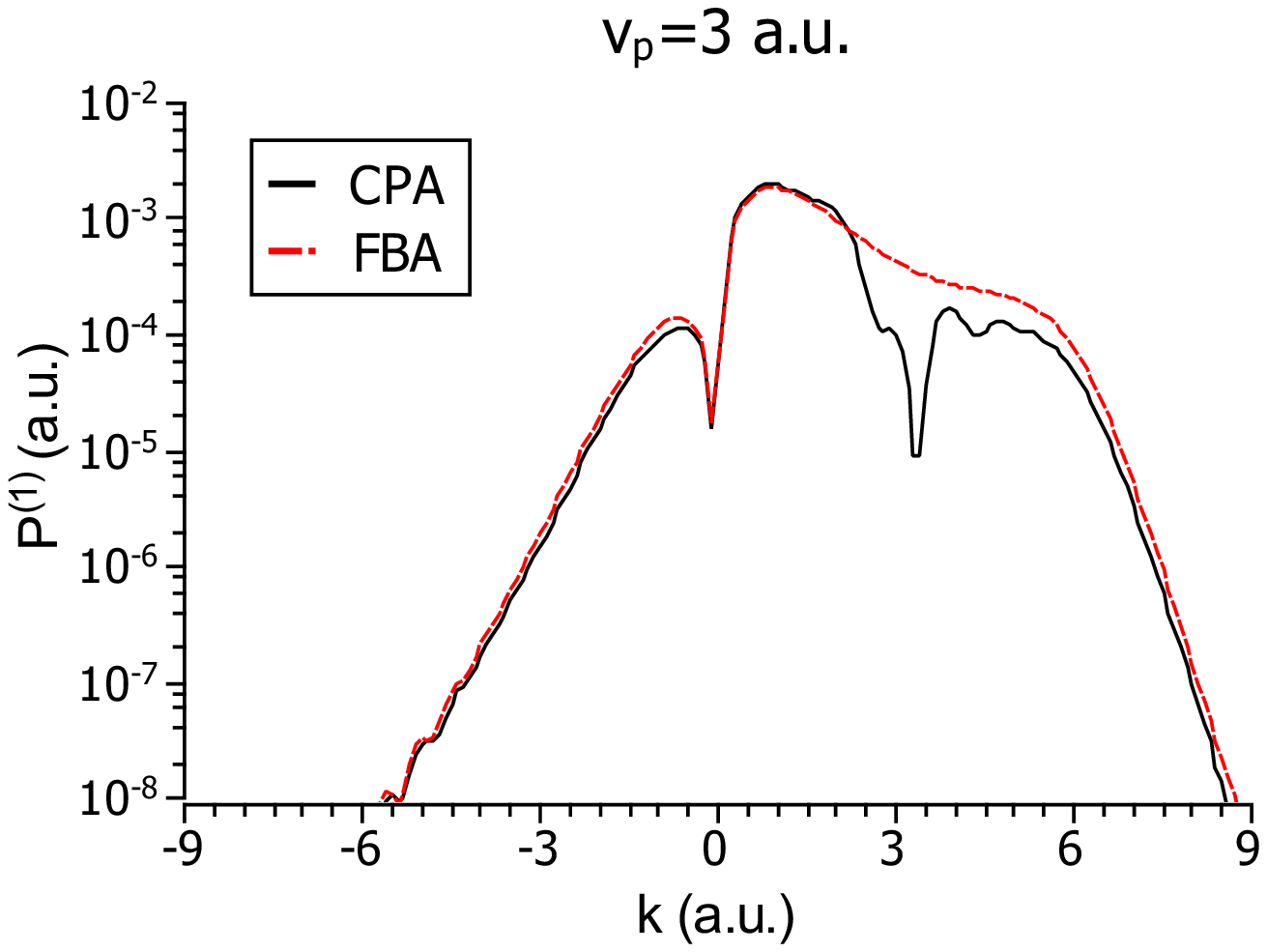}
\includegraphics[angle=-90,width=0.85\columnwidth]{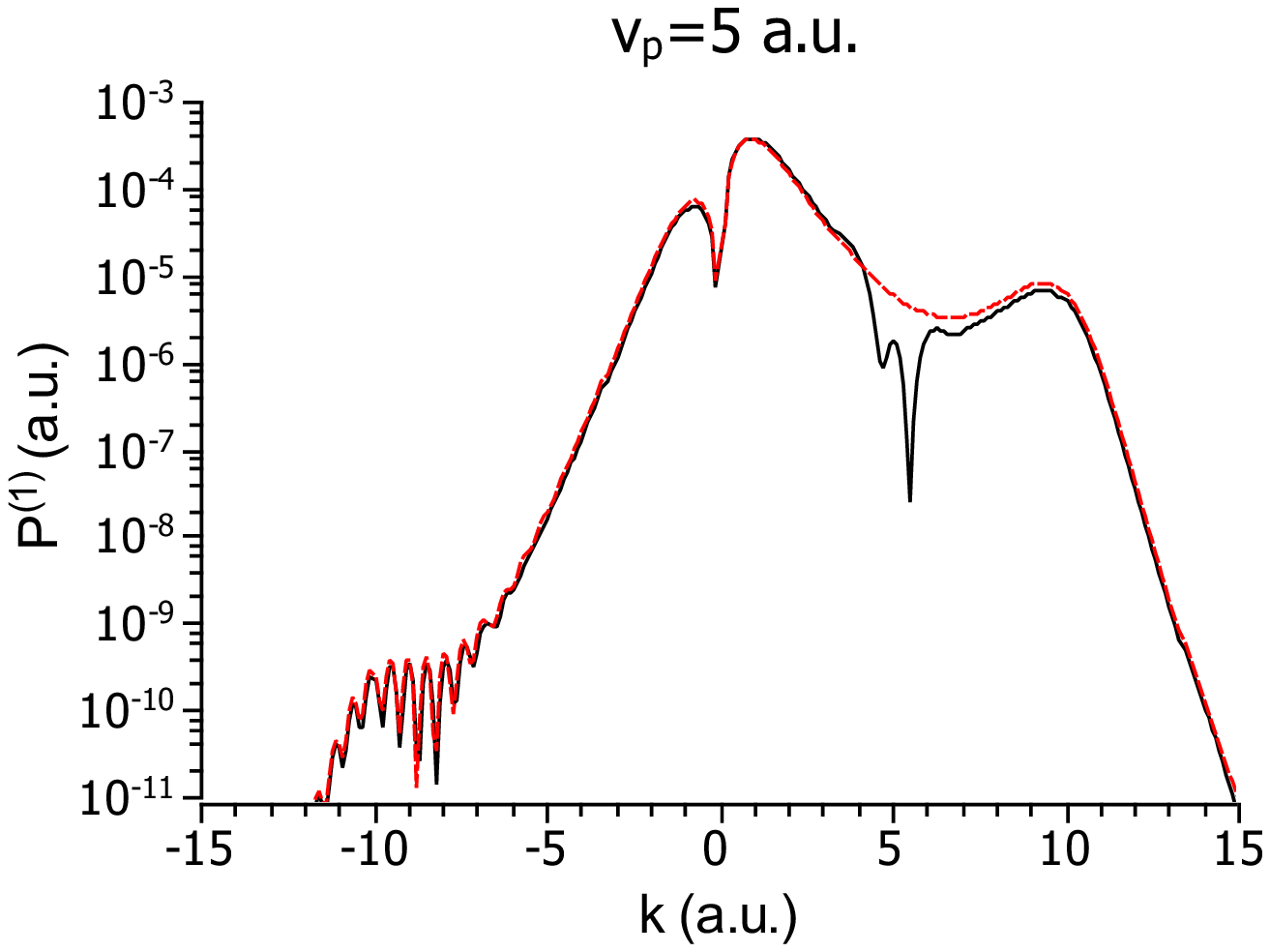}
\includegraphics[angle=-90,width=0.85\columnwidth]{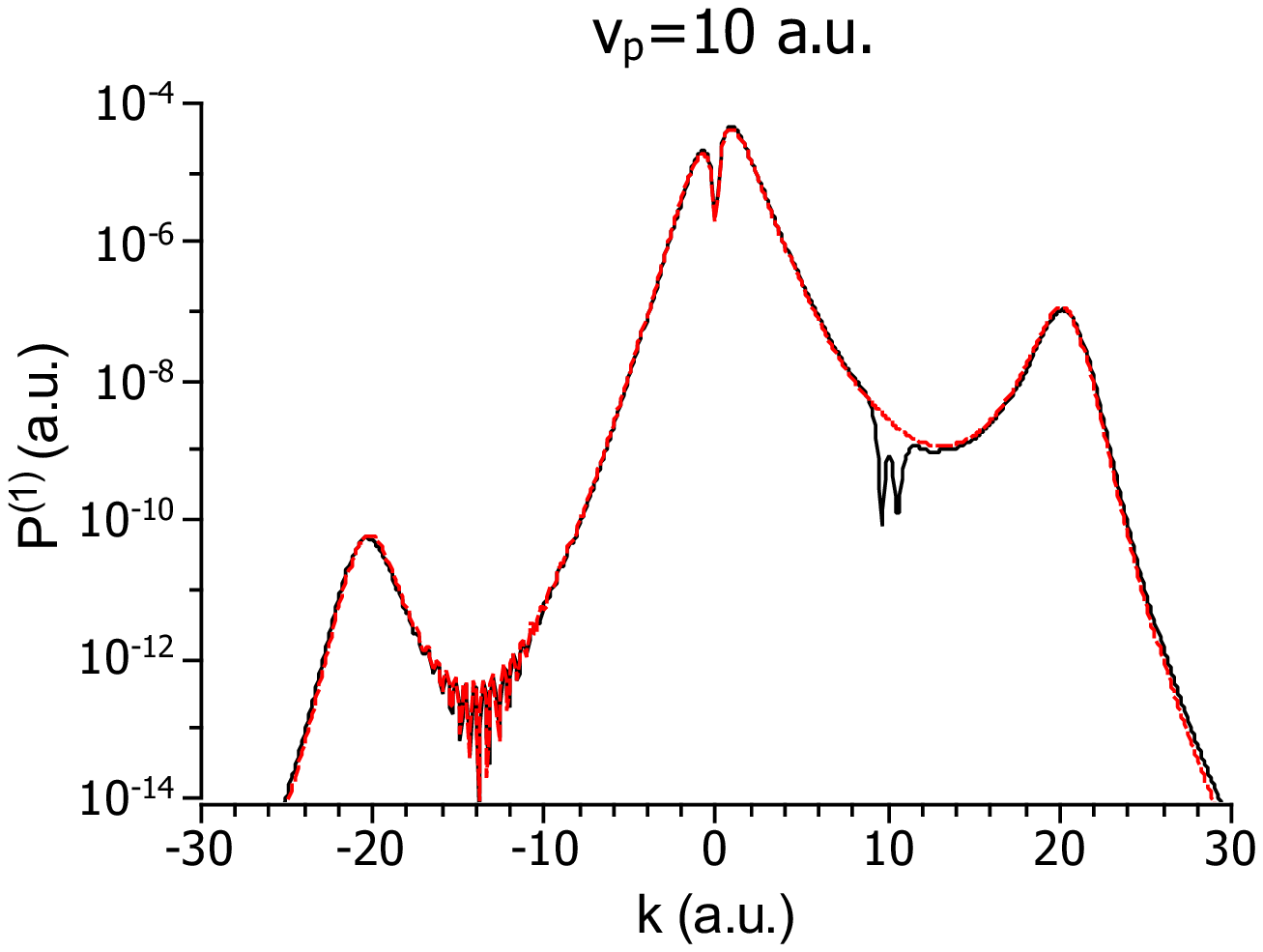}
\caption{(Color online) Ionization probability density $P^{(1)}(k)$ as
a function of the momentum of the ejected electron for the proton
velocities $v_p=3$~a.u. (top), 5~a.u. (middle) and
10~a.u. (bottom). The following calculations are shown: CPA (black
solid line) and FBA (red dashed line). }
\label{FIG_SI_vs_pe}
\end{figure} 

For further insight, we reduce TI to ionization and transfer of just
one electron. The FBA amplitude in this case can be expressed as  
\ba
 A_\text{SI1B}(k) &=& \la p_f k |U_1(x-x_p)| p_p
0 \ra \\ &&\hs{-2.5cm}=
 \frac{1}{\sqrt{v_pv_f}} \la k |
\exp\left(iqx\right) | 0 \ra
V_1(q)=\frac{1}{\sqrt{v_pv_f}}V_1(q)I_k(-q) \ ,
\nonumber 
\ea
where the momentum transfer is $q\simeq (k^2/2-\epsilon^2_0)/v_p$ 
and the initial and final state wave functions take the form
\ba
 \la x, x_p| p_p 0 \ra &=&
\frac{1}{\sqrt{v_p}}e^{ip_px_p}\varphi^2_0(x);\\ \la x, x_p| p_f k
\ra &=& \frac{1}{\sqrt{v_f}}e^{ip_fx_p}\varphi_{k}^{2(-)}(x).  
\ea
One can see from \Fref{FIG_SI_vs_pe} that the FBA and CPA results are
quite close except for the electron momenta $k\approx v_p$ where the
FBA overshoots CPA rather strongly. For these momenta, the electrons
are captured by the protons quite efficiently which is not accounted
by FBA.

\begin{figure}[ht]
\includegraphics[angle=-90,width=0.85\columnwidth]{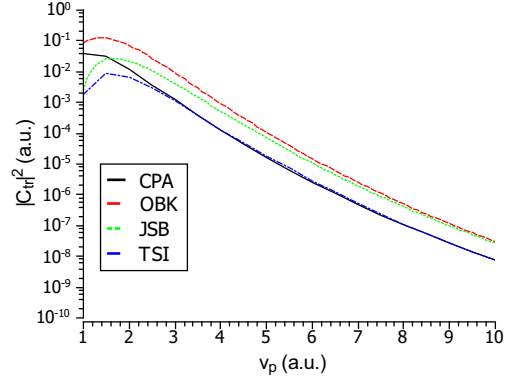}
\caption{(Color online) Transfer probability as a function of the proton
velocity $v_p$: CPA (black solid line), OBK (red dashed line), JSB (green dotted
line), TSI (blue dot-dashed line).}
\label{FIG_Ctr_vs_pe}
\end{figure} 

Now let us consider capture. 
The OBK amplitude in 1D kinematics has the form 
\ba
C_\text{OBK}&=&\la p_f f |U_1(x-x_p)| p_p 0 \ra \\
&=& \frac{1}{\sqrt{v_pv_H}} u^2_{0}(v_H-q) \int_{-\infty}^\infty
u^{1*}_0(\eta-q) V_1(\eta) d\eta 
\ ,
\nn
\ea
where the momentum transfer $q\simeq
(v_p^2/2+\epsilon^1_0-\epsilon^2_0)/v_p$, and the final state wave
function describes the hydrogen atom flying away with the velocity 
$v_H$
\ba \la x, x_p| p_H f \ra &=&
\frac{1}{\sqrt{v_H}}e^{ip_fx_p+iv_Hx}\varphi^1_{0}(x-x_p).  
\ea
It is seen from \Fref{FIG_Ctr_vs_pe} that the OBK
overshoots CPA quite considerably.
Now let us check if we can improve the OBK result by a simple
orthogonalization of the initial and final state wave functions.
\ba
 \la x, x_p| f_O \ra &=&
e^{iv_Hx}\varphi^{1}_{0}(x-x_p) \\
&-&
\la\varphi^2_0(x)|\,e^{iv_Hx}\varphi^{1}_{0}(x-x_p)\ra
\varphi^2_0(x) 
\nn
\ea

It is easy to show that the orthogonalization is equivalent to the
OBK formula with the original non-orthogonalized wave functions 
modified by a perturbation potential
\ba C_\text{JSB}&=&\la p_f f |U_1(x-x_p)-\bar{U}_1(x_p)| p_p 0
\ra \label{JSB} \ , \ea
where the balancing potential
\ba
\bar{U}_1(x_p)&=&\la 0 |U_1(x-x_p)| 0 \ra 
\ .
\ea
The balancing potential $\bar{U}_1(x_p)\approx U_1(x_p)$, where $-U_1(x_p)$ can be considered as proton-nucleus potential. Hence, for transfer from the neutral hydrogen, \Eref{JSB} is close to the JS approximation. 
Here we consider transfer from the helium ion, and a
formal application of the JS approximation for this case gives a
perturbation potential $U_1(x-x_p)-U_2(x_p)$. But Bates
\cite{Bates1952,Bates1958} and, lately, \citet{Lin1978A} have shown
that advantage of JS over OBK lie in the fact that the nuclear
potential compensates a spurious term from the non-orthogonality of
the initial and final state functions. The balancing potential should
be used instead of the nuclear potential in a general case \cite{Lin1978A}.
For this reasons, the approach given by \Eref{JSB} can be considered as
an improved JS approximation, and we named it as Jackson--Schiff--Bates
(JSB) approximation.
It is clear from \Fref{FIG_Ctr_vs_pe} that the OBK and JSB results
differ only at small $v_p$, but at large $v_p$ both calculations
significantly overshoot the CPA.

Now let us tackle the problem from another side. Suppose that, before
capture, a real ionization takes place and the ionized electron is
captured whose momentum distribution coincides with the momentum space
components of the electron bound to the proton. In this case, the TI
amplitude will be equal to the overlap integral
\ba
C_\text{trSI}&=&\int_{-\infty}^\infty u^{1*}_0(k-v_H)
A_\text{SI1B}(k) dk.  
\ea
We call this approach the transfer via single ionization (TSI) in the first
Born approximation. By inspecting \Fref{FIG_Ctr_vs_pe} one
can conclude that the TSI and CPA results are practically
coincident at large $v_p$.

In the time-dependent formalism, the TSI is equivalent to solving
the following TDSE
\ba
i\frac{\partial\psi_1(x,t)}{\partial
t}=\hat{h}_0\psi_1(x,t)+U_1(x-v_pt)\varphi^2_{0}(x)e^{-i\epsilon^2_{0}t}. \label{TDSE1B}
\ea
and subsequent projection of the solution $\psi_1(x,t)$ to the
function
\be
\tilde{\psi}_\text{tr}(x,t)=\int_{-\infty}^\infty u^{1}_0(k-v_H)
\varphi_{k}^{2(-)}(x) \exp\left(-i\frac{k^2}{2}t\right)
dk \label{ppsitr} 
\ee
instead of the function \eref{psitr}. Since the proton potential in
\Eref{TDSE1B} acts solely as a perturbation, the bound state of the
proton and electron cannot be described by this equation.  The
function \eref{ppsitr} describes the outgoing wavepacket, but unlike
\Eref{psitr} it is a solution of \Eref{TDSE1B} at $t\to\infty$. 
More broadly, the major difference of TSI and OBK/JS is that the
Born matrix element is calculated in the former between the
eigenfunctions of the same Hamiltonian $\hat{h}_0$, whereas in the
latter these functions belong to different Hamiltonians.

\subsection{Transfer ionization via double ionization} 

Similarly TSI for simple transfer, we develop a method to calculate TI
via double ionization (TIDI). The double ionization amplitude in
the FBA has the form
\be A_\text{DI1B}(k_1,k_2) =
\la p_Hk_1k_2 |U_1(x_1-x_p)+U_1(x_2-x_p)| p_p 0 \ra 
\ee
where the initial and final state wave functions
\ba
 \la x_1, x_2, x_p|
p_p 0 \ra &=& \frac{1}{\sqrt{v_p}}e^{ip_px_p}\varphi_0(x_1,x_2);\\
\la x_1, x_2, x_p| p_Hk_1k_2 \ra &=&
\frac{1}{\sqrt{v_f}}e^{ip_fx_p}\varphi_{k_1k_2}^{(-)}(x_1,x_2).  
\ea
After integration over $x_p$ we obtain
\be
A_\text{DI1B}(k_1,k_2) = \frac{1}{\sqrt{v_pv_f}} \la k_1k_2
|\exp\left(iqx_1\right)+\exp\left(iqx_2\right)| 0 \ra V_1(q), 
\ee
where the momentum transfer 
\be q=p_p-p_f\simeq
\frac{1}{v_p}\left(\frac{k_1^2}{2}+\frac{k_2^2}{2}-E_0\right).  
\ee
The TI amplitude is calculated as 
\ba
A_\text{trDI}(k)&=&\sqrt{2}\int_{-\infty}^\infty u^{1*}_0(k_1-v_H)
A_\text{DI1B}(k_1,k) dk_1.  
\ea
\begin{figure}[ht]
\includegraphics[angle=-90,width=0.85\columnwidth]{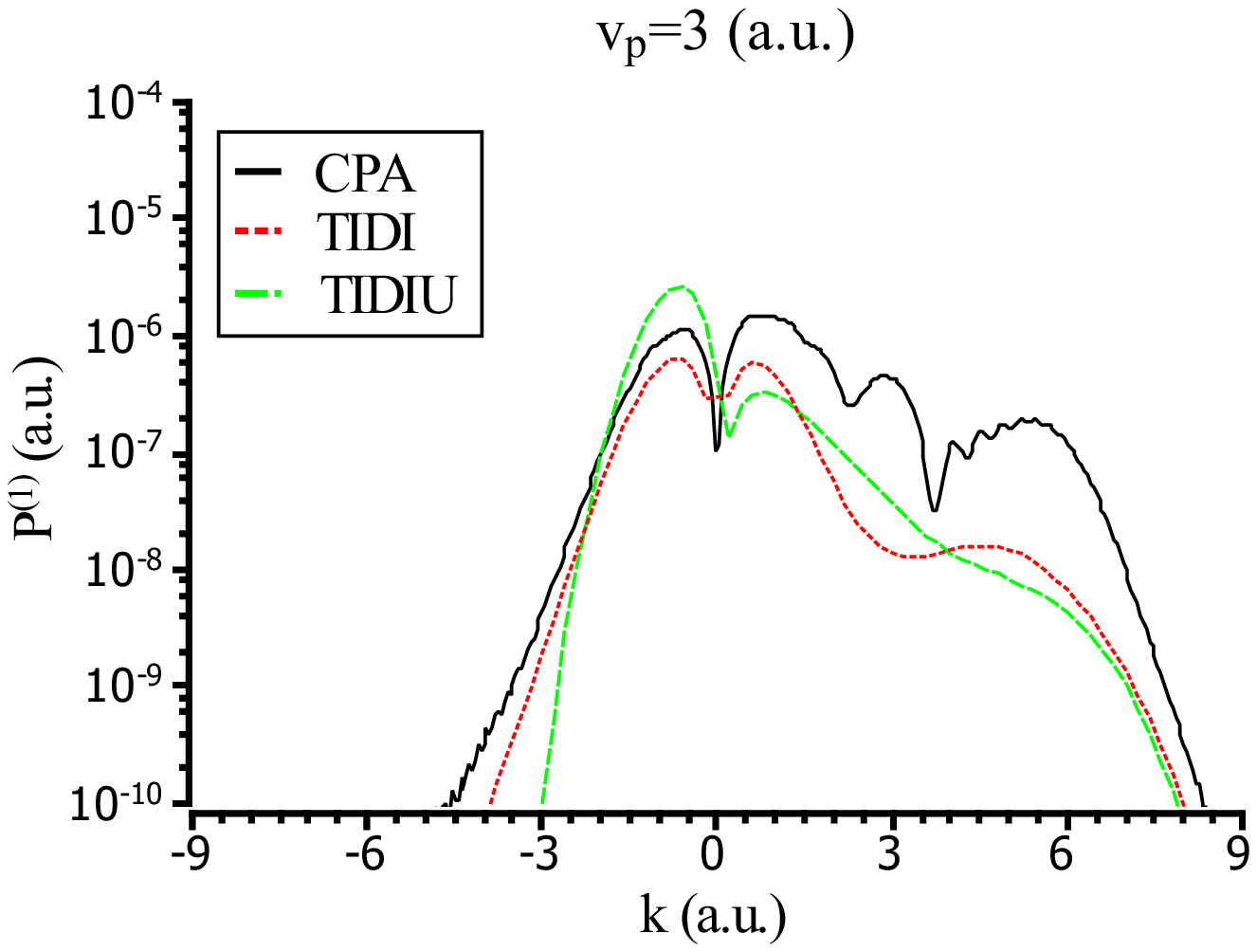}
\includegraphics[angle=-90,width=0.85\columnwidth]{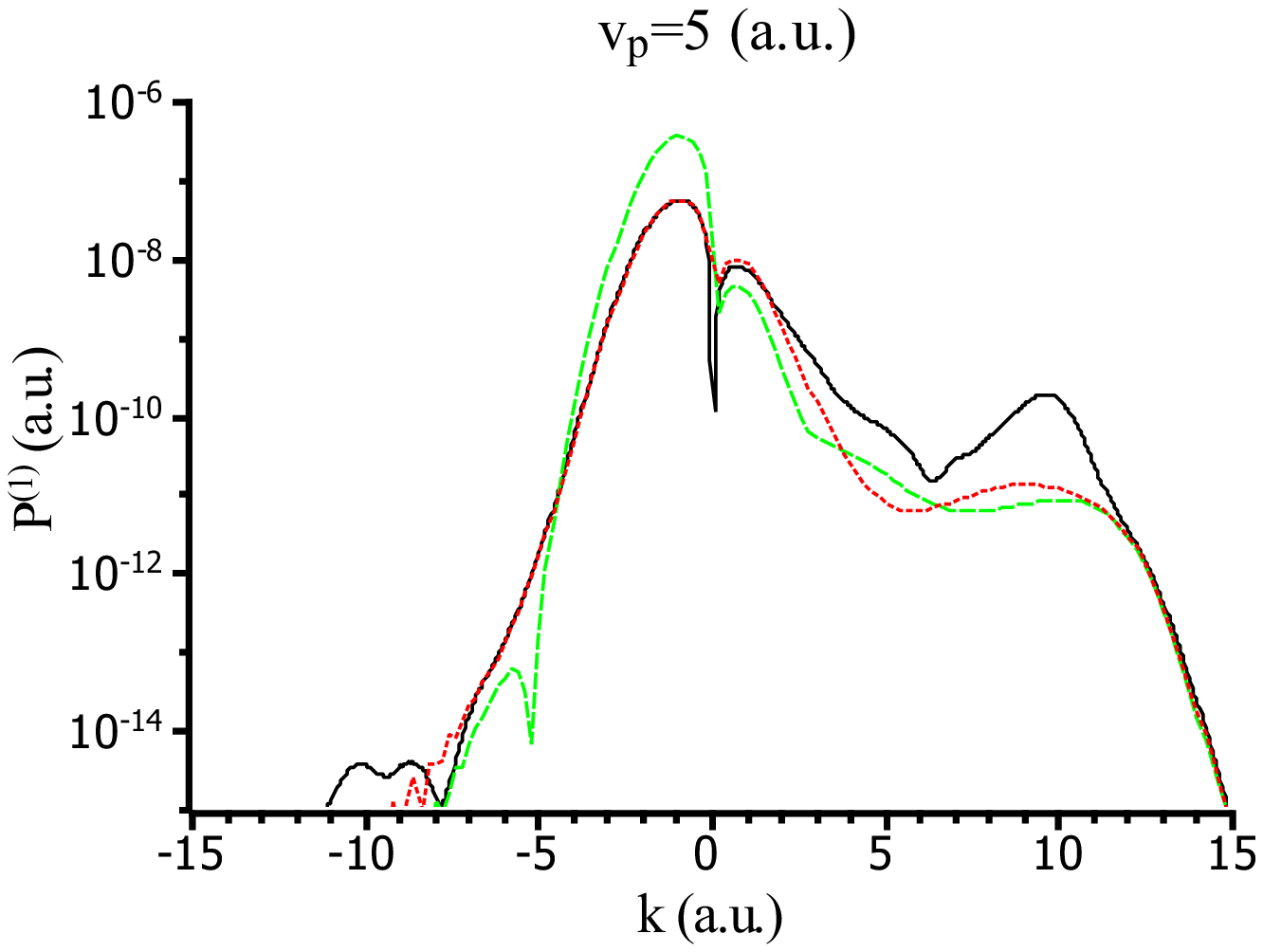}
\includegraphics[angle=-90,width=0.85\columnwidth]{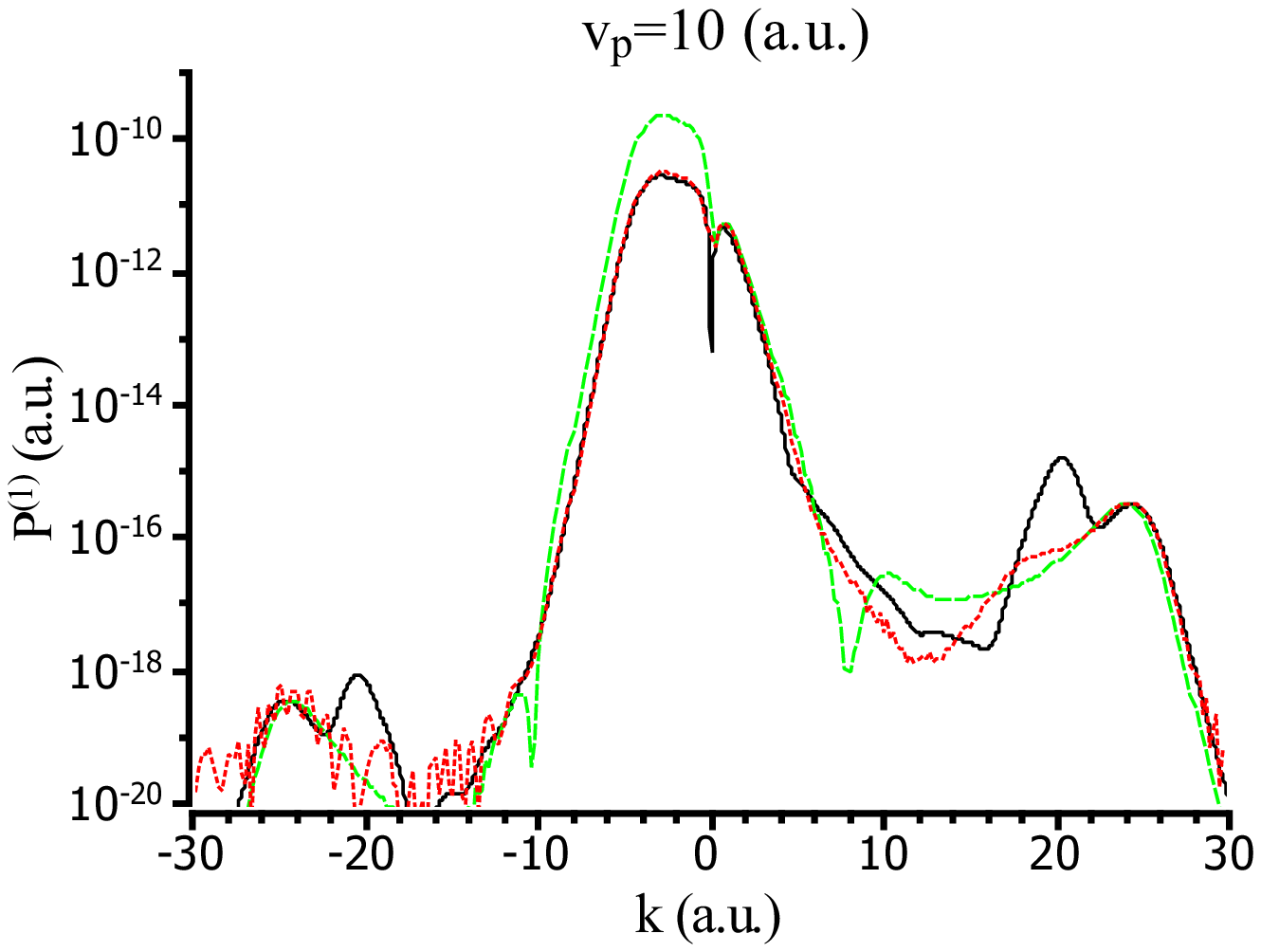}
\caption{(Color online) TI probability density $P^{(1)}(k)$ as
a function of the momentum of the transferred electron for the proton
velocities $v_p=3$~a.u. (top), 5~a.u. (middle) and
10~a.u. (bottom). Various calculations are shown as 
CPA (black solid line), TIDI (red dotted line), TIDIU
(green dashed line).}
\label{FIG_PDI_vs_pe}
\end{figure}

\begin{figure}[ht]
\includegraphics[angle=-90,width=0.85\columnwidth]{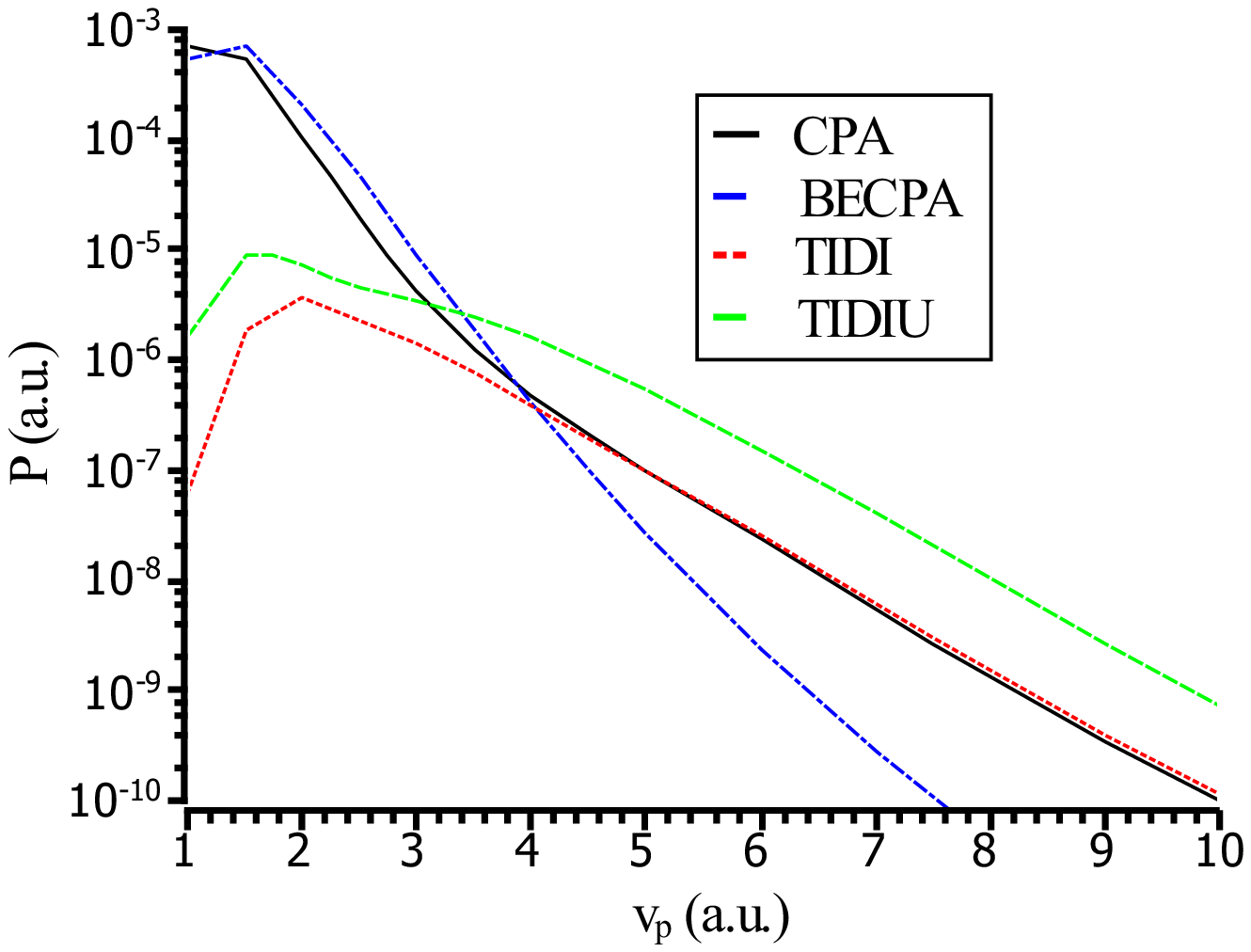}
\includegraphics[angle=-90,width=0.85\columnwidth]{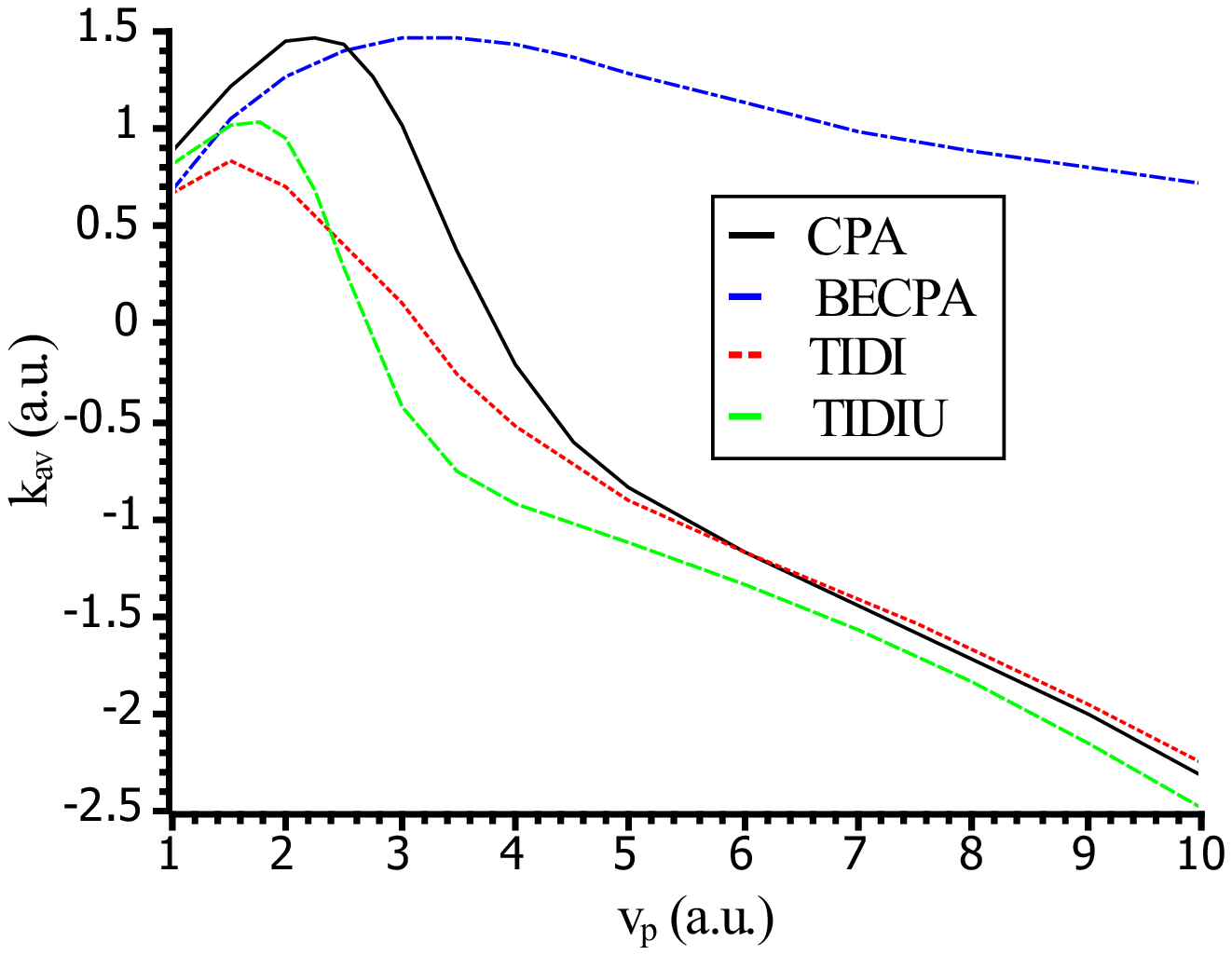}
\caption{(Color online) The total TI probability $P$ (top) and the
  mean momentum of ejected electron $\la k\ra$ (bottom) as
  functions of $v_p$. Various calculations are shown as CPA (black
  solid line), TIDI (red dotted line), TIDIU (green dashed
  line), BECPA (blue dash-dotted line).}
\label{FIG_kDI_vs_pe}
\end{figure} 

From \Fref{FIG_PDI_vs_pe} it is obvious that at large $v_p$ and small
$k$ the TIDI and CPA results are practically coincident.  And from
\Fref{FIG_kDI_vs_pe} it is clear that both the total TI probability
$P$ and the mean elected electron momentum $\la k\ra$, are
also very close between the TIDI and CPA in the whole region of
dominance of the SO. On the top panel of \Fref{FIG_kDI_vs_pe} the
cross-over between the BE and SO mechanisms is seen most graphically:
at $v_p<4$ the CPA curves is close to that of the BECPA, whereas at
$v_p>4$  to that of the TIDI.

From \Fref{FIG_PDI_vs_pe} one can observe that for large $k$ the
difference between TIDI and CPA is much more prominent. This is
explained by a large contribution of the BE process to TI at large $k$
even at large $v_p$. By comparing the bottom panels of
\Fref{FIG_P_vs_pe} and \Fref{FIG_PDI_vs_pe} one can discern some
interesting details explaining the origin of a double peak near
$k=2v_p$. As we mentioned before, this peak appears due to the
reflection of the ejected electrons by the proton potential.  It is
clearly seen on the bottom panel of \Fref{FIG_P_vs_pe} the  first
half of this peak (at lower momenta) is due to the BE process, and
from the bottom panel of \Fref{FIG_PDI_vs_pe} it is clear that the
second half (at larger momenta) is due to the SO. In the latter case,
the sequence events giving rise to TI is the following. First, the
impinging proton reflects the electron which flies away with the
velocity $k=2v_p$, then the second electron, emitted due to the SO, is
captured by the proton.

In 3D kinematics, TI can proceed via the nuclear or electron
Thomas processes \cite{Briggs1979}. In these processes, the electron
 scattered by the proton acquires a velocity $v_p$ or $\sqrt{2}v_p$
and then, in the secondary scattering from, respectively, the nucleus
or the second electron, receives a velocity $\v_p$ to be captured by the
proton. The Thomas processes are not possible in 1D kinematics because
the electron scattered on the proton acquires either zero or $2v_p$
velocity in this case.

To pinpoint the role of inter-electron interaction in the final state,
we also performed calculations with uncorrelated wave functions of the
two-electron continuum orthogonalized to the initial state 
\ba
\tilde{\varphi}_{k_1k_2}^{(-)}(x_1,x_2) = \chi_{k_1k_2}^{(-)}(x_1,x_2)
- \la\chi_{k_1k_2}^{(-)}|0\ra \varphi_0(x_1,x_2), 
\ea
where 
\ba
\chi_{k_1k_2}^{(-)}(x_1,x_2)
=\frac{1}{\sqrt{2}}\times
\\ &&\hs{-3cm}
\left[\varphi_{k_1}^{2(-)}(x_1)\varphi_{k_2}^{2(-)}(x_2)+\varphi_{k_2}^{2(-)}(x_1)\varphi_{k_1}^{2(-)}(x_2)\right].
\nn
\ea
This approximation is labeled as TIDIU.

From \Fref{FIG_kDI_vs_pe}, it is clear that the TIDIU overestimate
significantly the value of $P$. The mean momentum $\la k\ra$
is rather close to the exact solution, only slightly overestimated in
the negative region. It is seen in \Fref{FIG_PDI_vs_pe} that
overestimated $P$ is due to a strong growth of the peak at small
$k<0$, whereas at small $k>0$ the TIDIU calculation is close to
both the TIDI and CPA. Thus, we can conclude that the
inter-electron correlation in the final state suppresses partially the
backward emission and works out of sync with the correlation in the
initial state.

\section{Conclusion} 
We have performed time-dependent calculations of transfer ionization
in the fast proton scattering on the helium atom. We solved a
time-dependent Schr\"odinger equation under the classical projectile
motion approximation in one-dimensional kinematics. To gain deeper
physical insight into specific mechanisms of the TI reaction, we also
performed various perturbative 1D calculations which mimicked
realistic first Born calculations performed by other authors.  We
identified a strong effect of the non-orthogonality of the initial and
final states which may lead to some spurious unphysical terms.  This
term may be responsible for poor performance of the FBA employed by
other authors.

Among various approximate perturbative TI schemes, the most accurate
calculation is obtained by overlapping the double ionization amplitude
with the momentum profile of the final state wave function.  This
indicates that the most probable scenario of TI involves double
ionization and subsequent capture of those of the ionized electrons
which falls into the attractive potential well of the proton by
matching its velocity.

Because both Godunov {\em et al}
\cite{0953-4075-37-10-L01,PhysRevA.71.052712,PhysRevA.78.012714}, and
Popov {\em et al}
\cite{PhysRevA.81.032703,PhysRevA.87.032715,PhysRevA.88.042710} employ
a very similar formalism
based on the Jackson--Schiff approximation,
our remarks on possible flaws of the FBA concern both groups of
authors.  More broad is the question of using non-orthogonal initial
and final states belonging to different Hamiltonians in perturbative
calculations. This question goes beyond the scope of the present
work.
The same question was a matter of discussion in theory of simple
transfer much earlier \cite{Lin1978,Lin1978A,Bates1952,Bates1958}.

In future, we intend to extend our 1D model to full dimensionality
under the same classical projectile motion approximation.

We acknowledge Markus Sch\"offler and Reinhard D\"orner for critical
reading of the manuscript and Yuri Popov for useful and stimulating
discussions. 
V.V.S. acknowledges support from the Russian Foundation for Basic Research
(Grant No. 14-01-00420-a).

%


\begin{thebibliography}{18}
\expandafter\ifx\csname natexlab\endcsname\relax\def\natexlab#1{#1}\fi
\expandafter\ifx\csname bibnamefont\endcsname\relax
  \def\bibnamefont#1{#1}\fi
\expandafter\ifx\csname bibfnamefont\endcsname\relax
  \def\bibfnamefont#1{#1}\fi
\expandafter\ifx\csname citenamefont\endcsname\relax
  \def\citenamefont#1{#1}\fi
\expandafter\ifx\csname url\endcsname\relax
  \def\url#1{\texttt{#1}}\fi
\expandafter\ifx\csname urlprefix\endcsname\relax\def\urlprefix{URL }\fi
\providecommand{\bibinfo}[2]{#2}
\providecommand{\eprint}[2][]{\url{#2}}

\bibitem[{\citenamefont{Mergel {\em et~al}}(2001)\citenamefont{Mergel, D\"orner,
  Khayyat, Achler, Weber, Jagutzki, L\"udde, Cocke, and
  Schmidt-B\"ocking}}]{PhysRevLett.86.2257}
\bibinfo{author}{\bibfnamefont{V.}~\bibnamefont{Mergel}},
  \bibinfo{author}{\bibfnamefont{R.}~\bibnamefont{D\"orner}},
  \bibinfo{author}{\bibfnamefont{K.}~\bibnamefont{Khayyat}},
  \bibinfo{author}{\bibfnamefont{M.}~\bibnamefont{Achler}},
  \bibinfo{author}{\bibfnamefont{T.}~\bibnamefont{Weber}},
  \bibinfo{author}{\bibfnamefont{O.}~\bibnamefont{Jagutzki}},
  \bibinfo{author}{\bibfnamefont{H.~J.} \bibnamefont{L\"udde}},
  \bibinfo{author}{\bibfnamefont{C.~L.} \bibnamefont{Cocke}}, \bibnamefont{and}
  \bibinfo{author}{\bibfnamefont{H.}~\bibnamefont{Schmidt-B\"ocking}},
  \emph{\bibinfo{title}{Strong correlations in the {He} ground state momentum
  wave function observed in the fully differential momentum distributions for
  the $\mathit{p}+\mathrm{He}$ transfer ionization process}},
  \bibinfo{journal}{Phys. Rev. Lett.} \textbf{\bibinfo{volume}{86}},
  \bibinfo{pages}{2257} (\bibinfo{year}{2001}).

\bibitem[{\citenamefont{Schmidt-B\"ocking
  {\em et~al}}(2003)\citenamefont{Schmidt-B\"ocking, Mergel, D\"orner, Cocke,
  Jagutzki, Schmidt, Weber, Ludde, Weigold, Popov et~al.}}]{SB03}
\bibinfo{author}{\bibfnamefont{H.}~\bibnamefont{Schmidt-B\"ocking}},
  \bibinfo{author}{\bibfnamefont{V.}~\bibnamefont{Mergel}},
  \bibinfo{author}{\bibfnamefont{R.}~\bibnamefont{D\"orner}},
  \bibinfo{author}{\bibfnamefont{C.~L.} \bibnamefont{Cocke}},
  \bibinfo{author}{\bibfnamefont{O.}~\bibnamefont{Jagutzki}},
  \bibinfo{author}{\bibfnamefont{L.}~\bibnamefont{Schmidt}},
  \bibinfo{author}{\bibfnamefont{T.}~\bibnamefont{Weber}},
  \bibinfo{author}{\bibfnamefont{H.~J.} \bibnamefont{Ludde}},
  \bibinfo{author}{\bibfnamefont{E.}~\bibnamefont{Weigold}},
  \bibinfo{author}{\bibfnamefont{Y.~V.} \bibnamefont{Popov}},
  \bibnamefont{et~al.}, \emph{\bibinfo{title}{Revealing the non-$s^2$
  contributions in the momentum wave function of ground state {He}}},
  \bibinfo{journal}{Europhys. Lett.}
  \textbf{\bibinfo{volume}{62}}(\bibinfo{number}{4}), \bibinfo{pages}{477}
  (\bibinfo{year}{2003}).

\bibitem[{\citenamefont{Schmidt-B\"oking
  {\em et~al}}(2003)\citenamefont{Schmidt-B\"oking, Mergel, D\"orner, L\"udde,
  Schmidt, Weber, Weigold, and Kheifets}}]{Kheifets2003a}
\bibinfo{author}{\bibfnamefont{H.}~\bibnamefont{Schmidt-B\"oking}},
  \bibinfo{author}{\bibfnamefont{V.}~\bibnamefont{Mergel}},
  \bibinfo{author}{\bibfnamefont{R.}~\bibnamefont{D\"orner}},
  \bibinfo{author}{\bibfnamefont{H.~J.} \bibnamefont{L\"udde}},
  \bibinfo{author}{\bibfnamefont{L.}~\bibnamefont{Schmidt}},
  \bibinfo{author}{\bibfnamefont{T.}~\bibnamefont{Weber}},
  \bibinfo{author}{\bibfnamefont{E.}~\bibnamefont{Weigold}}, \bibnamefont{and}
  \bibinfo{author}{\bibfnamefont{A.~S.} \bibnamefont{Kheifets}},
  \emph{\bibinfo{title}{{Many-particle quantum dynamics in atomic and molecular
  fragmentation}}} (\bibinfo{publisher}{Springer},
  \bibinfo{address}{Heidelberg}, \bibinfo{year}{2003}), chap.
  \bibinfo{chapter}{Fast $p$-{He} transfer ionization processes: A window to
  reveal the non-$s^2$ contributions in the momentum wave function of ground
  state {He}}, pp. \bibinfo{pages}{353--378}.

\bibitem[{\citenamefont{Sch\"offler
  {\em et~al}}(2013{\natexlab{a}})\citenamefont{Sch\"offler, Chuluunbaatar, Popov,
  Houamer, Titze, Jahnke, Schmidt, Jagutzki, Galstyan, and
  Gusev}}]{PhysRevA.87.032715}
\bibinfo{author}{\bibfnamefont{M.~S.} \bibnamefont{Sch\"offler}},
  \bibinfo{author}{\bibfnamefont{O.}~\bibnamefont{Chuluunbaatar}},
  \bibinfo{author}{\bibfnamefont{Y.~V.} \bibnamefont{Popov}},
  \bibinfo{author}{\bibfnamefont{S.}~\bibnamefont{Houamer}},
  \bibinfo{author}{\bibfnamefont{J.}~\bibnamefont{Titze}},
  \bibinfo{author}{\bibfnamefont{T.}~\bibnamefont{Jahnke}},
  \bibinfo{author}{\bibfnamefont{L.~P.~H.} \bibnamefont{Schmidt}},
  \bibinfo{author}{\bibfnamefont{O.}~\bibnamefont{Jagutzki}},
  \bibinfo{author}{\bibfnamefont{A.~G.} \bibnamefont{Galstyan}},
  \bibnamefont{and} \bibinfo{author}{\bibfnamefont{A.~A.} \bibnamefont{Gusev}},
  \emph{\bibinfo{title}{Transfer ionization and its sensitivity to the
  ground-state wave function}}, \bibinfo{journal}{Phys. Rev. A}
  \textbf{\bibinfo{volume}{87}}, \bibinfo{pages}{032715}
  (\bibinfo{year}{2013}{\natexlab{a}}).

\bibitem[{\citenamefont{Sch\"offler
  {\em et~al}}(2013{\natexlab{b}})\citenamefont{Sch\"offler, Chuluunbaatar, Houamer,
  Galstyan, Titze, Schmidt, Jahnke, Schmidt-B\"ocking, D\"orner, Popov
  et~al.}}]{PhysRevA.88.042710}
\bibinfo{author}{\bibfnamefont{M.~S.} \bibnamefont{Sch\"offler}},
  \bibinfo{author}{\bibfnamefont{O.}~\bibnamefont{Chuluunbaatar}},
  \bibinfo{author}{\bibfnamefont{S.}~\bibnamefont{Houamer}},
  \bibinfo{author}{\bibfnamefont{A.}~\bibnamefont{Galstyan}},
  \bibinfo{author}{\bibfnamefont{J.~N.} \bibnamefont{Titze}},
  \bibinfo{author}{\bibfnamefont{L.~P.~H.} \bibnamefont{Schmidt}},
  \bibinfo{author}{\bibfnamefont{T.}~\bibnamefont{Jahnke}},
  \bibinfo{author}{\bibfnamefont{H.}~\bibnamefont{Schmidt-B\"ocking}},
  \bibinfo{author}{\bibfnamefont{R.}~\bibnamefont{D\"orner}},
  \bibinfo{author}{\bibfnamefont{Y.~V.} \bibnamefont{Popov}},
  \bibnamefont{et~al.}, \emph{\bibinfo{title}{Two-dimensional electron-momentum
  distributions for transfer ionization in fast proton-helium collisions}},
  \bibinfo{journal}{Phys. Rev. A} \textbf{\bibinfo{volume}{88}},
  \bibinfo{pages}{042710} (\bibinfo{year}{2013}{\natexlab{b}}).

\bibitem[{\citenamefont{Godunov {\em et~al}}(2005)\citenamefont{Godunov, Whelan,
  Walters, Schipakov, Sch\"offler, Mergel, D\"orner, Jagutzki, Schmidt, Titze
  et~al.}}]{PhysRevA.71.052712}
\bibinfo{author}{\bibfnamefont{A.~L.} \bibnamefont{Godunov}},
  \bibinfo{author}{\bibfnamefont{C.~T.} \bibnamefont{Whelan}},
  \bibinfo{author}{\bibfnamefont{H.~R.~J.} \bibnamefont{Walters}},
  \bibinfo{author}{\bibfnamefont{V.~S.} \bibnamefont{Schipakov}},
  \bibinfo{author}{\bibfnamefont{M.}~\bibnamefont{Sch\"offler}},
  \bibinfo{author}{\bibfnamefont{V.}~\bibnamefont{Mergel}},
  \bibinfo{author}{\bibfnamefont{R.}~\bibnamefont{D\"orner}},
  \bibinfo{author}{\bibfnamefont{O.}~\bibnamefont{Jagutzki}},
  \bibinfo{author}{\bibfnamefont{L.~P.~H.} \bibnamefont{Schmidt}},
  \bibinfo{author}{\bibfnamefont{J.}~\bibnamefont{Titze}},
  \bibnamefont{et~al.}, \emph{\bibinfo{title}{Transfer ionization process
  $p+\mathrm{He}\rightarrow{\mathrm{H}}^{0}+{\mathrm{He}}^{2+}+{e}^{-}$ with
  the ejected electron detected in the plane perpendicular to the incident beam
  direction}}, \bibinfo{journal}{Phys. Rev. A} \textbf{\bibinfo{volume}{71}},
  \bibinfo{pages}{052712} (\bibinfo{year}{2005}).

\bibitem[{\citenamefont{Godunov {\em et~al}}(2004)\citenamefont{Godunov, Whelan, and
  Walters}}]{0953-4075-37-10-L01}
\bibinfo{author}{\bibfnamefont{A.~L.} \bibnamefont{Godunov}},
  \bibinfo{author}{\bibfnamefont{C.~T.} \bibnamefont{Whelan}},
  \bibnamefont{and} \bibinfo{author}{\bibfnamefont{H.~R.~J.}
  \bibnamefont{Walters}}, \emph{\bibinfo{title}{Fully differential cross
  sections for transfer ionization --- a sensitive probe of high level correlation
  effects in atoms}}, \bibinfo{journal}{J.~Phys.~B}
  \textbf{\bibinfo{volume}{37}}(\bibinfo{number}{10}), \bibinfo{pages}{L201}
  (\bibinfo{year}{2004}).
	
\bibitem[{\citenamefont{Godunov {\em et~al}}(2008)\citenamefont{Godunov, Whelan, and
  Walters}}]{PhysRevA.78.012714}
\bibinfo{author}{\bibfnamefont{A.~L.} \bibnamefont{Godunov}},
  \bibinfo{author}{\bibfnamefont{C.~T.} \bibnamefont{Whelan}},
  \bibnamefont{and} \bibinfo{author}{\bibfnamefont{H.~R.~J.}
  \bibnamefont{Walters}}, \emph{\bibinfo{title}{Effect of angular electron
  correlation in He: Second-order calculations for transfer ionization}},
  \bibinfo{journal}{Phys. Rev. A} \textbf{\bibinfo{volume}{78}},
  \bibinfo{pages}{012714} (\bibinfo{year}{2008}).

\bibitem[{\citenamefont{Houamer {\em et~al}}(2010)\citenamefont{Houamer, Popov, and
  Dal~Cappello}}]{PhysRevA.81.032703}
\bibinfo{author}{\bibfnamefont{S.}~\bibnamefont{Houamer}},
  \bibinfo{author}{\bibfnamefont{Y.~V.} \bibnamefont{Popov}}, \bibnamefont{and}
  \bibinfo{author}{\bibfnamefont{C.}~\bibnamefont{Dal~Cappello}},
  \emph{\bibinfo{title}{Failure of the multiple peaking approximation for fast
  capture processes at milliradian scattering angles}}, \bibinfo{journal}{Phys.
  Rev. A} \textbf{\bibinfo{volume}{81}}, \bibinfo{pages}{032703}
  (\bibinfo{year}{2010}).

\bibitem[{\citenamefont{Kheifets {\em et~al}}(1999)\citenamefont{Kheifets, Bray, and
  Bartschat}}]{0953-4075-32-15-104}
\bibinfo{author}{\bibfnamefont{A.~S.} \bibnamefont{Kheifets}},
  \bibinfo{author}{\bibfnamefont{I.}~\bibnamefont{Bray}}, \bibnamefont{and}
  \bibinfo{author}{\bibfnamefont{K.}~\bibnamefont{Bartschat}},
  \emph{\bibinfo{title}{Convergent calculations for simultaneous
  electron-impact ionization-excitation of helium}},
  \bibinfo{journal}{J.~Phys.~B}
  \textbf{\bibinfo{volume}{32}}(\bibinfo{number}{15}), \bibinfo{pages}{L433}
  (\bibinfo{year}{1999}).

\bibitem[{\citenamefont{Voitkiv {\em et~al}}(2008)\citenamefont{Voitkiv, Najjari and
  Ullrich}}]{PhysRevLett.101.223201}
\bibinfo{author}{\bibfnamefont{A.~B.} \bibnamefont{Voitkiv}},
\bibinfo{author}{\bibfnamefont{B.} \bibnamefont{Najjari}},
  \bibnamefont{and}
  \bibinfo{author}{\bibfnamefont{J.}~\bibnamefont{Ullrich}},
  \emph{\bibinfo{title}{Mechanism for electron transfer in fast ion-atomic
  collisions}}, \bibinfo{journal}{Phys. Rev. Lett.}
  \textbf{\bibinfo{volume}{101}}, \bibinfo{pages}{223201}
  (\bibinfo{year}{2008}).

\bibitem[{\citenamefont{Voitkiv}(2008)}]{0953-4075-41-19-195201}
\bibinfo{author}{\bibfnamefont{A.~B.} \bibnamefont{Voitkiv}},
  \emph{\bibinfo{title}{Electron-electron interaction and transfer ionization
  in fast ion-atom collisions}}, \bibinfo{journal}{J.~Phys.~B}
  \textbf{\bibinfo{volume}{41}}(\bibinfo{number}{19}), \bibinfo{pages}{195201}
  (\bibinfo{year}{2008}).

\bibitem[{\citenamefont{Voitkiv and Ma}(2012)}]{PhysRevA.86.012709}
\bibinfo{author}{\bibfnamefont{A.~B.} \bibnamefont{Voitkiv}} \bibnamefont{and}
  \bibinfo{author}{\bibfnamefont{X.}~\bibnamefont{Ma}},
  \emph{\bibinfo{title}{Dynamics of transfer ionization in fast ion-atom
  collisions}}, \bibinfo{journal}{Phys. Rev. A} \textbf{\bibinfo{volume}{86}},
  \bibinfo{pages}{012709} (\bibinfo{year}{2012}).

\bibitem[{\citenamefont{Popov {\em et~al}}(2014)\citenamefont{Popov, Shablov,
  Kouzakov, and Galstyan}}]{PhysRevA.89.036701}
\bibinfo{author}{\bibfnamefont{Y.~V.} \bibnamefont{Popov}},
  \bibinfo{author}{\bibfnamefont{V.~L.} \bibnamefont{Shablov}},
  \bibinfo{author}{\bibfnamefont{K.~A.} \bibnamefont{Kouzakov}},
  \bibnamefont{and} \bibinfo{author}{\bibfnamefont{A.~G.}
  \bibnamefont{Galstyan}}, \emph{\bibinfo{title}{Comment on ``dynamics of
  transfer ionization in fast ion-atom collisions''}}, \bibinfo{journal}{Phys.
  Rev. A} \textbf{\bibinfo{volume}{89}}, \bibinfo{pages}{036701}
  (\bibinfo{year}{2014}).

\bibitem[{\citenamefont{Voitkiv and Ma}(2014)}]{PhysRevA.89.036702}
\bibinfo{author}{\bibfnamefont{A.~B.} \bibnamefont{Voitkiv}} \bibnamefont{and}
  \bibinfo{author}{\bibfnamefont{X.}~\bibnamefont{Ma}},
  \emph{\bibinfo{title}{Reply to ``comment on `dynamics of transfer ionization
  in fast ion-atom collisions' ''}}, \bibinfo{journal}{Phys. Rev. A}
  \textbf{\bibinfo{volume}{89}}, \bibinfo{pages}{036702}
  (\bibinfo{year}{2014}).

\bibitem[{\citenamefont{McGuire}(1997)}]{McGuire1997}
\bibinfo{author}{\bibfnamefont{J.~H.} \bibnamefont{McGuire}},
  \emph{\bibinfo{title}{Electron Correlation Dynamics in Atomic Collisions}},
  Cambridge Monographs on Atomic, Molecular and Chemical Physics
  (\bibinfo{publisher}{Cambridge University Press},
  \bibinfo{address}{Cambridge}, \bibinfo{year}{1997}).

\bibitem[{\citenamefont{Serov {\em et~al}}(2013)\citenamefont{Serov, Derbov,
  Sergeeva, and Vinitsky}}]{PhysRevA.88.043403}
\bibinfo{author}{\bibfnamefont{V.~V.} \bibnamefont{Serov}},
  \bibinfo{author}{\bibfnamefont{V.~L.} \bibnamefont{Derbov}},
  \bibinfo{author}{\bibfnamefont{T.~A.} \bibnamefont{Sergeeva}},
  \bibnamefont{and} \bibinfo{author}{\bibfnamefont{S.~I.}
  \bibnamefont{Vinitsky}}, \emph{\bibinfo{title}{Hybrid surface-flux method for
  extraction of the ionization amplitude from the calculated wave function}},
  \bibinfo{journal}{Phys. Rev. A} \textbf{\bibinfo{volume}{88}},
  \bibinfo{pages}{043403} (\bibinfo{year}{2013}).

\bibitem[{\citenamefont{Landau and Lifshitz}(1985)}]{LL85}
\bibinfo{author}{\bibfnamefont{L.~D.} \bibnamefont{Landau}} \bibnamefont{and}
  \bibinfo{author}{\bibfnamefont{E.~M.} \bibnamefont{Lifshitz}},
  \emph{\bibinfo{title}{Quantum Mechanics (Non-relativistic theory)}},
  vol.~\bibinfo{volume}{3} of \emph{\bibinfo{series}{Course of theoretical
  physics}} (\bibinfo{publisher}{Pergamon press, Oxford},
  \bibinfo{year}{1985}), \bibinfo{edition}{3rd} ed.

\bibitem[{\citenamefont{Bates}(1952)}]{Bates1952}
\bibinfo{author}{\bibfnamefont{D.~R.} \bibnamefont{Bates}},
\bibinfo{author}{\bibfnamefont{A.} \bibnamefont{Dalgarno}}, 
  \emph{\bibinfo{title}{Electron Capture I: Resonance Capture from Hydrogen Atoms by Fast Protons}},
  \bibinfo{journal}{Proc. Phys. Soc. A} \textbf{\bibinfo{volume}{65}},
  \bibinfo{pages}{919} (\bibinfo{year}{1952}).

\bibitem[{\citenamefont{Bates}(1958)}]{Bates1958}
\bibinfo{author}{\bibfnamefont{D.~R.} \bibnamefont{Bates}}, 
  \emph{\bibinfo{title}{Electron Capture in Fast Collisions}},
  \bibinfo{journal}{Proc. R. Soc. Lond. A} \textbf{\bibinfo{volume}{247}},
  \bibinfo{pages}{294} (\bibinfo{year}{1958}).
	
\bibitem[{\citenamefont{Lin {\em et~al}}(1978)}]{Lin1978A}
\bibinfo{author}{\bibfnamefont{C.~D.} \bibnamefont{Lin}}, 
\bibinfo{author}{\bibfnamefont{S.~C.} \bibnamefont{Soong}}, 
\bibinfo{author}{\bibfnamefont{L.~N.} \bibnamefont{Tunnell}}, 
  \emph{\bibinfo{title}{Two-state atomic expansion methods for electron capture from multielectron atoms
by fast protons}},
  \bibinfo{journal}{Phys. Rev. A} \textbf{\bibinfo{volume}{17}},
  \bibinfo{pages}{1646} (\bibinfo{year}{1978}).

\bibitem[{\citenamefont{Briggs and Taulbjerg}(1979)}]{Briggs1979}
\bibinfo{author}{\bibfnamefont{J.~S.} \bibnamefont{Briggs}},
  \bibinfo{author}{\bibfnamefont{K.} \bibnamefont{Taulbjerg}},
  \emph{\bibinfo{title}{Charge transfer by a double-scattering mechanism involving 
target electrons}},
  \bibinfo{journal}{J. Phys. B: Atom. Molec. Phys.} \textbf{\bibinfo{volume}{12}},
  \bibinfo{pages}{2565} (\bibinfo{year}{1979}).

\bibitem[{\citenamefont{Lin}(1978)}]{Lin1978}
\bibinfo{author}{\bibfnamefont{C.~D.} \bibnamefont{Lin}}, 
  \emph{\bibinfo{title}{Electron capture for ion-atom collisions at intermediate energies}},
  \bibinfo{journal}{J. Phys. B} \textbf{\bibinfo{volume}{11}},
  \bibinfo{pages}{L185} (\bibinfo{year}{1978}).

\end{thebibliography}

\end{document}